\documentclass[%
 reprint,
 amsmath,amssymb,
 aps,
]{revtex4-1}

\usepackage[version=3]{mhchem} 

\usepackage{balance} 
\usepackage{rotating}
\usepackage{url}
\usepackage{tabularx}
\usepackage{multirow}
\usepackage{amsmath}
\usepackage{color, colortbl}
\usepackage{bm}
\usepackage[table]{xcolor}
\usepackage{dcolumn}
\usepackage{longtable}
\usepackage{nicefrac}
\usepackage{tablefootnote}
\usepackage{threeparttable}
\usepackage{makecell}
\AtBeginDocument{}
\usepackage{chemmacros}
\newcommand{\Lower}[1]{\smash{\lower 1.5ex \hbox{#1}}}

\newcolumntype{d}[1]{D{.}{.}{#1}}

\begin{document}
\author{Katharina Boguslawski}
\email{k.boguslawski@fizyka.umk.pl}
\affiliation{Institute of Physics, Faculty of Physics, Astronomy and Informatics, Nicolaus Copernicus University in Torun, Grudziadzka 5, 87-100 Torun, Poland}
\affiliation{Departement of Chemistry, Nicolaus Copernicus University in Torun, Gagarina 7, 87-100 Torun, Poland}
\author{Pawe{\l} Tecmer}
\email{ptecmer@fizyka.umk.pl}
\affiliation{Institute of Physics, Faculty of Physics, Astronomy and Informatics, Nicolaus Copernicus University in Torun, Grudziadzka 5, 87-100 Torun, Poland}

\title[Benchmark of dynamic electron correlation models for seniority-zero wavefunctions and their application to thermochemistry]
 {Benchmark of dynamic electron correlation models for seniority-zero wavefunctions and their application to thermochemistry}
 
 
\begin{abstract}
Wavefunctions restricted to electron-pair states are promising models to describe static/nondynamic electron correlation effects encountered, for instance, in bond-dissociation processes and transition-metal and actinide chemistry.
To reach spectroscopic accuracy, however, the missing dynamic electron correlation effects that cannot be described by electron-pair states need to be included \textit{a posteriori}.
In this article, we extend the previously presented perturbation theory models with an Antisymmetric Product of 1-reference orbital Geminal (AP1roG) reference function that allow us to describe both static/nondynamic and dynamic electron correlation effects.
Specifically, our perturbation theory models combine a diagonal and off-diagonal zero-order Hamiltonian, a single-reference and multi-reference dual state, and different excitation operators used to construct the projection manifold.
We benchmark all proposed models as well as an \textit{a posteriori} linearized coupled cluster correction on top of AP1roG against CR-CCSD(T) reference 
data for reaction energies of several closed-shell molecules that are extrapolated to the basis set limit. 
Moreover, we test the performance of our new methods for multiple bond breaking processes in the N$_2$, C$_2$, and BN dimers against MRCI-SD and MRCI-SD+Q reference data.      
Our numerical results indicate that the best performance is obtained from a linearized coupled cluster correction as well as second-order perturbation theory corrections employing a diagonal and off-diagonal zero-order Hamiltonian and a single-determinant dual state.
These dynamic corrections on top of AP1roG allow us to reliably model molecular systems dominated by static/nondynamic as well as dynamic electron correlation.
        
\end{abstract}
\maketitle

\section{Introduction}
Accurate and reliable theoretical predictions of thermodynamic properties of chemical reactions remain an active field of research~\cite{peterson2012chemical}. 
Computational studies are particularly desirable especially when experimental manipulations are limited or impossible. 
Conventional ``ab initio'' quantum chemistry tools used in thermochemistry are commonly based on different flavours of many-body perturbation theory and coupled cluster approaches~\cite{Coester_1958,Cizek_jcp_1966,Cizek_Paldus_1971,Paldus_Cizek_Shavitt_1972,Bartlett_rev_1981,Helgaker_book,Shavitt_book,bartlett_2007}. 
Over the years, the CCSD(T) (Coupled Cluster Singles, Doubles and perturbative Triples) method has unfolded as one of the most reliable and accurate models for systems that are well-represented by a single electron configuration. 
Therefore, it is often referred to as the ``gold standard'' of quantum chemistry. 
Although CCSD(T) performs extraordinary well for systems dominated by dynamic correlation, the model often fails for multi-reference systems. 
For molecular systems with a substantial multi-reference character, multi-reference methods~\cite{Szalay2012} are commonly used, which are computationally very expensive.
Furthermore, the most popular multi-reference approaches used in quantum chemistry do not~\textit{a priori} account for the dynamic part of the correlation energy. 
Exceptions are multi-reference coupled cluster models, some variants of the multi-reference configuration interaction method, and the perturb-then-diagonalize multi-reference perturbation theory approaches that use effective Hamiltonians. 
To remedy this problem, \textit{a posteriori} corrections have been developed that are usually based on perturbation theory~\cite{caspt21,caspt22,nevpt2,Pulay_CASPT2_2011,RASPT2}, canonical transformation theory~\cite{CTT-review} or coupled cluster methods~\cite{monika_mrcc}. 

Although multi-reference methods are frequently applied to model strongly-correlated systems, like bond-breaking processes and heavy-element chemistry, some flavours of single-reference coupled cluster theory pose promising alternatives to describe static/nondynamic electron correlation.
Among the most representative examples are the CR-CC (Completely Renormalized Coupled Cluster) and CR-EOMCC (Completely Renormalized Equation of Motion Coupled Cluster) approaches~\cite{CR-CCa,CR-CCb,CR-CCc,CR-CCd,CR-EOMCCSD}, the active-space CC and EOMCC methods~\cite{kowalski2000active,kowalski2005active,kowalski2010active,piecuch2010active}, the EA/IP- (Electron Affinity/Ionization Potential) and DEA/DIP-(Double EA/IP) EOMCC models~\cite{2-2-EOM-theory}, and the spin-flip CC/EOMCC formalism~\cite{spin-flip-EOM,spin-flip-EOM-theory,krylov2008equation}. 
A different, computationally feasible approach suitable for strongly-correlated systems uses seniority-zero wavefunctions to describe the static/nondynamic part of the electron correlation energy.~\cite{OO-AP1roG,pawel_jpca_2014,PS2-AP1roG,AP1roG-JCTC,pawel_PCCP2015,Rassolov2002,Surjan2012,Neuscamman_2012,Limacher_2013,Ellis2013,Piotrus_Mol-Phys,Tamar-pCC,p-CCD,Eric-orbital-optimization} 
The missing dynamic electron correlation effects are included~\textit{a posteriori} in these ans{\"a}tze using, for instance, many-body perturbation theory~\cite{Jeszenszki2014,Piotrus_PT2,Limacher_2015}, coupled-cluster theory~\cite{Zoboki2013,Gordon-VB-CC,frozen-pCCD,Kasia-lcc,dmrg-cc}, extended random phase approximation~\cite{Pernal2014}, and density functional theory (DFT) corrections~\cite{Garza2015,Garza-pccp}. 

In this article, we seek computationally cheap and reliable dynamic electron correlation models on top of the Antisymmetric Product of 1-reference orbital Geminal (AP1roG) of different levels of approximation.
The AP1roG wavefunction can be written using an exponential ansatz,
\begin{equation}\label{eq:ap1rog}
|{\rm AP1roG}\rangle = \exp \left (  \sum_{i=1}^P \sum_{a=P+1}^K c_i^a a_a^{\dagger}  a_{\bar{a}}^{\dagger}a_{\bar{i}} a_{i}  \right )| 0 \rangle,
\end{equation}
where the sum runs over all electron pairs $P$ (equal to the number of occupied orbitals) and virtual orbitals $K-P$, while $\vert 0 \rangle$ is some reference determinant.
$a_p^\dagger$ and $a_p$ are the fermionic creation and annihilation operators for spin-up $p$ and spin-down $\bar{p}$ electrons, while $c_i^a $ are the AP1roG amplitudes, also called geminal coefficients.
The geminal coefficients $c_i^a $ are obtained by solving the projected Schr\"odinger equation, as in coupled cluster theory.
The AP1roG model allows us to describe static/nondynamic correlation and to a minor extent some dynamic correlation (see also ref.~\cite{Boguslawski2016}).
To include the missing part of the dynamic electron correlation effects \textit{a posteriori} that go beyond the AP1roG ansatz, that is, beyond electron pair states, we will use perturbation theory as well as a linearized coupled cluster corrections.~\cite{Kasia-lcc}
Specifically, our perturbation theory models will combine a diagonal and off-diagonal zero-order Hamiltonian, a single- and multi-determinant wavefunction as dual, and different excitation operators used to construct the projection manifold. 
The performance of all perturbation theory models as well as the linearized coupled cluster correction, as presented in ref.~\citenum{Kasia-lcc}, will be assessed against spectroscopic constants of homo- and heteronuclear dimers and reaction energies of closed shell systems. 

This article is organized as follows. In section~\ref{sec:theory}, we present different theoretical models to correct for the missing dynamic electron correlation in the AP1roG ansatz.
Computational details are described in section~\ref{sec:compdetails}.
Numerical results and comparison to standard electron correlation methods for molecular systems dominated by static/nondynamic (bond dissociation processes) and dynamic electron correlation (reaction energies of main group compounds) are presented in section~\ref{sec:results}.
Finally, we conclude in section~\ref{sec:conclusion}.
\section{Theoretical models for dynamic correlation}\label{sec:theory}
A common way to account for dynamic correlation effects in quantum many-body systems is to use perturbation theory (PT).
Specifically for the AP1roG wavefunction, two different PT models have been proposed that allow us to describe electron correlation effects beyond electron pairs.~\cite{Piotrus_PT2}
While these approaches provide reliable spectroscopic constants for some first-row diatomic molecules,~\cite{pawel_jpca_2014,Kasia-lcc} their performance deteriorates when moving to heavy-element containing compounds like actinide species.~\cite{Kasia-lcc}
To remedy this problem, a linearized coupled cluster (LCC) corrections on top of AP1roG can be used that allows us to accurately describe closed-shell molecules with (quasi-)degenerate $f$-, $d$-, $p$-, and $s$-shells, as encountered in actinide-containing compounds.~\cite{Kasia-lcc}
In the following, we will briefly summarize some theoretical models for dynamic correlation with an AP1roG reference function.
Specifically, we will extend the existing PT models and relate the proposed models to the recently presented PTa and PTb methods.~\cite{Piotrus_PT2}

\subsection{2nd-order Perturbation Theory}
One drawback of multi-reference perturbation theory is the arbitrariness of the theoretical model.
For example, there are different choices for the zero-order Hamiltonian $\hat{H}_0$, the dual state $\langle \tilde{\Psi} \vert$ in the projector, and the choice of the projection space.~\cite{Pulay_CASPT2_2011}
Poor choices may lead to technical difficulties and unphysical solutions.
In this work, we will investigate different selections for the zero-order Hamiltonian $\hat{H}_0$, the dual state $\langle \tilde{\Psi} \vert$, and the projection space, while the zero-order wavefunction is restricted to the AP1roG reference function of eq.~\eqref{eq:ap1rog}, $\vert \Psi^{(0)} \rangle = \vert \mathrm{AP1roG} \rangle$.

Furthermore, in the derivation of all PT models, we will use the quantum chemical Hamiltonian in its normal product from \textit{shifted} by the correlation energy of AP1roG $E_{\mathrm{corr}}^{(0)}$ (the shift in energy is indicated by ``$^\prime$''),
        \begin{align}\label{eq:nph}
            \hat{H}_N^\prime &= \hat{H} - \langle 0 \vert \hat{H} \vert 0 \rangle - E_{\mathrm{corr}}^{(0)}\nonumber \\
                        &= (\hat{H}_0 - \langle 0 \vert \hat{H}_0 \vert 0 \rangle) + (\hat{V} - \langle 0 \vert \hat{V} \vert 0 \rangle - E_{\mathrm{corr}}^{(0)} ) \nonumber \\
                        &= (\hat{H}_0)_N + \hat{V}_N^\prime,
        \end{align}
where the quantum chemical Hamiltonian $\hat{H}$ is divided into a zero-order contribution $\hat{H}_0$ and a perturbation $\hat{V}$.
It is convenient to rewrite $\hat{H}$ into a sum of a one- ($\hat{H}_1$) and a two-electron ($\hat{H}_2$, here indicated by $\hat{W}$) part,
        \begin{equation}
            \hat{H} = \hat{H}_1 + \hat{W} = \sum_{pq} h_{pq} a^\dagger_p a_q + \frac{1}{2} \sum_{pqrs} \langle pq \vert rs\rangle a^\dagger_p a^\dagger_q a_s a_r.
        \end{equation}
In the above equation, $h_{pq}$ and $\langle pq \vert rs\rangle$ are the one- and two-electron (written in physicists' notation) integrals, respectively, determined for the one-particle basis functions $p,q,r,s$.
We should note that, in this work, we will restrict $\hat{H}_0$ to be a one-body operator so that the normal-product form of the perturbation $\hat{V} = \hat{H}-\hat{H}_0$ can be written as an operator shifted by the AP1roG correlation energy, $\hat{V}_N^\prime = \hat{V}_N - E_{\mathrm{corr}}^{(0)}$ (again indicated by ``$^\prime$''), as we have $E_{\mathrm{corr}}^{(0)} = \langle 0 \vert \hat{W}_N \vert \mathrm{AP1roG} \rangle$.
Introducing a shifted perturbation operator $\hat{V}^\prime_N$ is equivalent to neglecting contractions (or diagrams) that correspond to the AP1roG correlation energy in the PT equations, which will be indicated by the ${}^\prime$ in the sum of the $\hat{W}_N^\prime$ operator.

As in conventional Rayleigh--Schr\"odinger PT, the exact wavefunction can be written as an order-by-order expansion, $\vert \Psi \rangle = \vert \mathrm{AP1roG} \rangle + \lambda \vert \Psi^{(1)} \rangle + \ldots$, where $\lambda$ is the order parameter.
The first-order correction to the wavefunction is expanded in a set of Slater determinants $\Phi_p$,
        \begin{equation}\label{eq:psi1}
            \vert {\Psi^{(1)}} \rangle = \sum_p t_p \vert \Phi_p \rangle,
        \end{equation}
and forced to be orthogonal to the zero-order wavefunction, here, $\vert \mathrm{AP1roG} \rangle$,
        \begin{equation}\label{eq:orthopt}
            \langle {\Psi^{(1)}} \vert \mathrm{AP1roG} \rangle = 0.
        \end{equation}
This orthogonality constraint restricts the choice of the projection space used for the expansion of $ \vert {\Psi^{(1)}} \rangle$ (and higher orders).
By construction, all pair-excitations with respect to $ \vert 0 \rangle$ have to be excluded as they don't satisfy $\langle \Phi_p\vert \mathrm{AP1roG} \rangle = 0$.
Similarly, introducing an order parameter $\lambda$ in the Hamiltonian $\hat{H} = \hat{H}_0 + \lambda \hat{V}$ and equating coefficients of powers of $\lambda$, we obtain the zero-, first-, and higher-order PT equations. Specifically for the first-order correction to the wavefunction, we have to solve
        \begin{align}
            (\hat{H}_0)_N \vert \Psi^{(1)} \rangle + \hat{V}_N^\prime \vert \mathrm{AP1roG} \rangle &= 0 \nonumber \\
            \sum_p t_p (\hat{H}_0)_N \vert \Phi_p \rangle + \hat{V}_N^\prime \vert \mathrm{AP1roG} \rangle &= 0.
        \end{align}
Since we have introduced a shifted normal-product Hamiltonian, the zero- and first-order energy corrections vanish,
        \begin{equation}\label{eq:e01}
            E^{(0)}+E^{(1)} = \frac{1}{\langle \tilde{\Psi}\vert \mathrm{AP1roG}\rangle} \langle \tilde{\Psi} \vert \hat{V}_N^\prime \vert \mathrm{AP1roG} \rangle = 0,
        \end{equation}
where $\langle \tilde{\Psi}\vert$ is the dual of the unperturbed state $\vert \mathrm{AP1roG}\rangle$. Specific choices for $\langle \tilde{\Psi}\vert$ will be considered below.
The first non-zero correction to the energy is of second-order and can be calculated from the first-order wavefunction and the shifted normal-product perturbation Hamiltonian,
        \begin{equation}\label{eq:e2}
            E^{(2)} = \frac{ \langle \tilde{\Psi} \vert \hat{V}_N^\prime \vert \Psi^{(1)} \rangle }{\langle \tilde{\Psi}\vert \mathrm{AP1roG}\rangle}.
        \end{equation}
Before we focus on possible choices of $(\hat{H}_0)_N$ and $\hat{V}_N^\prime$ as well as $\langle \tilde{\Psi} \vert$, we will define our projection space used in the expansion of $\vert \Psi^{(1)} \rangle $ in eq.~\eqref{eq:psi1}.
Following previous PT models as well as linear CC corrections with an AP1roG reference function, the projection space will contain all possible excitations with respect to a reference determinant.
This reference determinant is, however, not arbitrary, but restricted to the reference determined of AP1roG, $\vert 0 \rangle$ of eq.~\eqref{eq:ap1rog}.
We should emphasize that $\vert 0 \rangle$ is not equivalent to the Hartree--Fock determinant as in conventional CC theory, but adjusted during the optimization of the AP1roG wavefunction.
Choosing $\vert 0 \rangle$ as reference determinant, the first-order correction can be written as
        \begin{equation}
            \vert \Psi^{(1)} \rangle = \hat{T} \vert 0 \rangle,
        \end{equation}
where $\hat{T}$ is some excitation operator that substitutes electrons from the occupied to the virtual space with respect to $\vert 0 \rangle$.
Furthermore, we will restrict $\hat{T}$ to contain double excitations without electron pairs, $\hat{T} = \hat{T}_2^\prime$, as well as single and double excitations, $\hat{T} = \hat{T}_1 + \hat{T}_2^\prime$.
If only double excitations are included, the excitation operator is specified as
        \begin{equation}\label{eq:t2}
            \hat{T}_2^\prime = \frac{1}{2} \sum_{ij}^{\rm occ}\sum_{ab}^{\rm virt}{^\prime} t_{ij}^{ab} \hat{E}_{ai} \hat{E}_{bj},
        \end{equation}
where $\hat{E}_{ai} = a^\dagger_{a}a_i+a^\dagger_{\bar{a}}a_{\bar{i}}$ is the singlet excitation operator and the perturbation amplitudes $t_{ij}^{ab}$ are symmetric with respect to pair-exchange, \emph{i.e.}, $t_{ij}^{ab}=t_{ji}^{ba}$.
Similar to our notation of the shifted normal-product Hamiltonian, the prime in the above summation indicates that pair-excited determinants are excluded in the excitation operator, \emph{i.e.}, $t_{i\bar{i}}^{a\bar{a}} = 0$.
Exclusion of pair-excitations fulfils the orthogonality condition eq.~\eqref{eq:orthopt}.
If the excitation operator contains both single and double excitations, the single excitations can be accounted for by adding
        \begin{equation}\label{eq:t1}
            \hat{T}_1 = \sum_{i}^{\rm occ}\sum_{a}^{\rm virt} t_{i}^{a} \hat{E}_{ai}
        \end{equation}
to the double excitation operator. 
Furthermore, as basis for the bra states of the projection manifold, we follow refs.~\citenum{Piotrus_PT2,Kasia-lcc} and use the convenient choice
        \begin{equation}
            \langle \overline{^{ab}_{ij}} \vert = \frac{1}{3} \langle ^{ab}_{ij} \vert + \frac{1}{6} \langle ^{ab}_{ji} \vert,
        \end{equation}
for doubly excited determinants where $\langle ^{ab}_{ij} \vert = \langle 0 \vert \hat{E}_{jb}\hat{E}_{ia}$ and
        \begin{equation}
            \langle \overline{^{a}_{i}} \vert = \frac{1}{2} \langle ^{a}_{i} \vert ,
        \end{equation}
for singly excited determinants with $\langle ^{a}_{i} \vert = \langle 0 \vert \hat{E}_{ia}$. 
The bra basis then forms a biorthogonal basis that satisfies the normalization conditions
        \begin{equation}
            \langle \overline{^{ab}_{ij}} \vert ^{cd}_{kl} \rangle = \delta_{iajb,kcld} + \delta_{jbia,kcld}
        \end{equation}
for doubly-excited determinants, while for the singly-excited determinants we have
        \begin{equation}
            \langle \overline{^{a}_{i}} \vert ^{b}_{j} \rangle = \delta_{ia,jb}.
        \end{equation}
Choosing eq.~\eqref{eq:psi1} as the first-order correction to the wavefunction with excitation operators as defined in eqs.~\eqref{eq:t2} and \eqref{eq:t1}, the corresponding perturbation amplitudes are then obtained by solving
        \begin{equation}\label{eq:pt}
            \sum_p t_p \langle \overline{\Phi_q} \vert (\hat{H}_0)_N \vert \Phi_p \rangle + \langle \overline{\Phi_q} \vert \hat{V}_N^\prime \vert \mathrm{AP1roG} \rangle = 0,
        \end{equation}
Note that the above equations depend on the partition of the Hamiltonian $\hat{H}$ into the zero-order and the perturbation part as well as the projection manifold, but not on the choice of the dual state $\langle \tilde{\Psi} \vert$.
In the following, we will consider different partitionings of $\hat{H}$ as well as two choices for $\langle \tilde{\Psi} \vert$.

\subsubsection{Diagonal one-body zero-order Hamiltonian}
First, we will restrict $(\hat{H}_0)_N$ to be a diagonal one-electron operator. Specifically, we will choose $(\hat{H}_0)_N$ to be the diagonal of the inactive Fock operator 
        \begin{align}\label{eq:fock}
            \hat{F} &= \sum_{pq} \left(h_{pq} + \sum_i^{\rm occ} ( \langle pi||qi \rangle + \langle pi | qi\rangle)\right) a^\dagger_p a_q \nonumber \\
                    &= \sum_{pq} f_{pq} a^\dagger_p a_q,
        \end{align}
where $\langle pi||qi \rangle $ are the two-electron integrals in physicists' notation containing the Coulomb $\langle pi|qi \rangle $ and exchange $\langle pi|iq \rangle $ terms.
Using normal-product operators, the zero-order Hamiltonian reads 
        \begin{equation}\label{eq:h0d}
            (\hat{H}_0)_N  = \hat{F}_N^{\mathrm{d}} = \sum_{p} f_{pp} \{a^\dagger_p a_p\},
        \end{equation}
while the perturbation becomes (cf.~eq.~\eqref{eq:nph})
        \begin{align}\label{eq:vd}
            \hat{V}_N^\prime  &= \hat{F}_N^{\mathrm{o}} + \hat{W}_N^\prime \nonumber \\
                              &= \sum_{p\neq q} f_{pq} \{a^\dagger_p a_q\} + \frac{1}{2}\sum_{pqrs}{}^\prime  \langle pq \vert rs\rangle \{ a^\dagger_p  a^\dagger_q a_s a_r \}
        \end{align}
Eqs.~\eqref{eq:h0d} and \eqref{eq:vd} are then substituted into eq.~\eqref{eq:pt} to solve for the PT amplitudes.
Note that, in the case of a diagonal zero-order Hamiltonian, the PT amplitudes are obtained from a set of uncoupled equations.

\paragraph{Restricting the dual $\langle \tilde{\Psi} \vert$ to $\langle 0\vert$.}
If $\langle \tilde{\Psi} \vert$ is restricted to the AP1roG reference determinant $\langle 0\vert$, we can straightforwardly evaluate the overlap
$\langle \Phi_0\vert \mathrm{AP1roG} \rangle$, which equals 1 due to intermediate normalization of the AP1roG wavefunction.
The expression for the second-order energy correction $E^{(2)}$ given in eq.~\eqref{eq:e2} thus simplifies to
        \begin{equation}\label{eq:epta}
            E^{(2)} = \langle 0  \vert \hat{V}_N^\prime \vert \Psi^{(1)} \rangle.
        \end{equation}
Specifically, the energy correction for $\hat{T}= \hat{T}^\prime_2$ is given as
        \begin{equation}\label{eq:e2sdd}
           E^{(2)}_{\rm d} = \sum_{iajb} t_{ij}^{ab} ( \langle ij \vert\vert ab \rangle + \langle ij \vert ab \rangle),
        \end{equation}
while for single and double excitations we have
        \begin{equation}\label{eq:e2sdsd}
           E^{(2)}_{\rm sd} = 2\sum_{ia}f_{ia}t_i^a + \sum_{iajb} t_{ij}^{ab} ( \langle ij \vert\vert ab \rangle + \langle ij \vert ab \rangle).
        \end{equation}
We should emphasize that this PT model is equivalent to the PTa model presented in ref.~\citenum{Piotrus_PT2}.
Note, however, that pair excitations are not excluded in ref.~\citenum{Piotrus_PT2} and that the full Hamiltonian $\hat{H}$ is taken as the perturbation Hamiltonian so that $E^{(2)} = \langle 0  \vert \hat{H} \vert \Psi^{(1)} \rangle$.
Since the PTa amplitude equations for the pair excitations vanish (as they equal the AP1roG amplitude equations), the PTa energy corrections is nonetheless determined from eq.~\eqref{eq:epta}.
For reasons of consistency, we will abbreviate these PT models using a single determinant (SD) as dual and a diagonal (d) zero-order Hamiltonian as PT2SDd.
The choice of the excitation operator will be indicated in parentheses, that is, PT2SDd(d) for double excitations and PT2SDd(sd) for both single and double excitations.

\paragraph{Choosing $\langle \mathrm{AP1roG} \vert$ as dual $\langle \tilde{\Psi} \vert$.}
If $\langle \mathrm{AP1roG} \vert$ is taken as the dual state $\langle \tilde{\Psi} \vert$, we have to evaluate terms as $\langle \mathrm{AP1roG} \vert\mathrm{AP1roG}\rangle$ in the energy expression, which becomes computationally intractable for large systems.
In order to arrive at a computationally feasible model, we will follow ref.~\citenum{Piotrus_PT2} to, at least partially, eliminate the overlap $\langle \mathrm{AP1roG} \vert \mathrm{AP1roG} \rangle$ in the PT equations and energy expression.
For that purpose, we redefine the zero-order Hamiltonian of eq.~\eqref{eq:h0d} by introducing the inverse of the overlap $\langle \mathrm{AP1roG} \vert \mathrm{AP1roG} \rangle$ as a scaling factor,
        \begin{align}\label{eq:h0dmd}
            (\hat{H}_0)_N &= \bar{{F}}_N^{\mathrm{d}} = \sum_{p} \frac{f_{pp}}{\langle \mathrm{AP1roG} \vert \mathrm{AP1roG} \rangle} \{a^\dagger_p a_p\} \nonumber \\
                          &= \sum_{p} \bar{f}_{pp} \{a^\dagger_p a_p\}.
        \end{align}
By changing the zero-order Hamiltonian of eq.~\eqref{eq:h0d}, we also have to adjust the corresponding perturbation part given in eq.~\eqref{eq:vd},
        \begin{align}\label{eq:vdmd}
            \hat{V}_N^\prime  &= \hat{F}_N-\bar{{F}}_N^{\mathrm{d}} + \hat{W}_N^\prime \nonumber \\
                              &= \sum_{p,q} (f_{pq}-\bar{f}_{pp}\delta_{pq}) \{a^\dagger_p a_q\} + \frac{1}{2}\sum_{pqrs}{}^\prime  \langle pq \vert rs\rangle \{ a^\dagger_p  a^\dagger_q a_s a_r \}.
        \end{align}
To fully avoid the evaluation of the overlap $\langle \mathrm{AP1roG} \vert \mathrm{AP1roG} \rangle$ in the PT amplitude equations eq.~\eqref{eq:pt}, the inverse of the AP1roG wavefunction overlap will be absorbed in the PT amplitudes. Thus, the first-order wavefunction contains scaled amplitudes, 
        \begin{equation}\label{eq:psi1md}
            \vert \overline{\Psi^{(1)}} \rangle = \sum_p \frac{t_p}{\langle \mathrm{AP1roG} \vert \mathrm{AP1roG} \rangle} \vert \Phi_p \rangle =  \sum_p \bar{t}_p \vert \Phi_p \rangle.
        \end{equation}
By substituting eq.~\eqref{eq:h0dmd} into eq.~\eqref{eq:pt} and introducing the scaled PT amplitudes from eq.~\eqref{eq:psi1md}, we can eliminate the wavefunction overlap from the PT working equations.
Furthermore, scaling the PT amplitudes by the inverse of the overlap $\langle \mathrm{AP1roG} \vert \mathrm{AP1roG} \rangle$ restores the zero-order Hamiltonian of eq.~\eqref{eq:h0d} and we get $(\hat{H}_0)_N  = \hat{{F}}_N^{\mathrm{d}}$.
Note that the wavefunction overlap is still present in the perturbation part $\hat{V}_N^\prime $.
Due to the structure of the zero-order wavefunction and the choice of the projection manifold, the diagonal part of the modified Fock operator in eq.~\eqref{eq:vdmd} does not contribute to the PT amplitude equations and the resulting perturbation reduces to $\hat{V}_N^\prime  = \hat{{F}}_N^{\mathrm{o}} + \hat{W}_N^\prime$.
Since we have chosen $\langle \mathrm{AP1roG} \vert$ as dual, the second-order energy correction is determined from $E^{(2)} = \langle \mathrm{AP1roG}  \vert \hat{V}_N^\prime \vert \overline{\Psi^{(1)}} \rangle$ (cf.~eq.~\eqref{eq:e2}), where the first-order correction to the wavefunction is calculated from the scaled PT amplitudes (eq.~\eqref{eq:psi1md}),
        \begin{equation}\label{eq:e2md}
            E^{(2)} = \sum_p \bar{t}_p \langle \mathrm{AP1roG}  \vert \hat{V}_N^\prime \vert \Phi_p \rangle.
        \end{equation}
The sum in the above equation runs over all determinants in the projection manifold (doubly excited or singly- and doubly-excited determinants).
We should note that although we can exactly evaluate the energy correction and the PT equations, the zero- and first-order energy corrections eq.~\eqref{eq:e01} do not vanish.
However, we can neglect the weights of the (AP1roG) amplitudes beyond single pair excitations and assume that $E^{(0)}+E^{(1)} \approx 0$.~\cite{Piotrus_PT2}
We will label this PT model as PT2MDd as it uses a multi-determinant (MD) wavefunction as dual and a diagonal (d) zero-order Hamiltonian.
Furthermore, PT2MDd(d) indicates that the excitation operator contains only double excitations (without pairs), while in PT2MDd(sd) both single and double excitations are included in $\hat{T}$.

\subsubsection{Off-diagonal one-body zero-order Hamiltonian}
Next, we will consider an off-diagonal one-electron operator as the zero-order Hamiltonian.
Similar to the PT models employing a diagonal zero-order Hamiltonian, we choose the Fock operator of eq.~\eqref{eq:fock} as our $\hat{H}_0$ Hamiltonian,
        \begin{equation}\label{eq:h0o}
            (\hat{H}_0)_N  = \hat{F}_N^{\mathrm{d}} + \hat{F}_N^{\mathrm{o}} = \sum_{p,q} f_{pq} \{a^\dagger_p a_q\}.
        \end{equation}
Then, the perturbation part $\hat{V}=\hat{H}-\hat{H}_0$ in its (shifted) normal-product form reads
        \begin{equation}\label{eq:vo}
            \hat{V}_N^\prime  = \frac{1}{2}\sum_{pqrs}{}^\prime  \langle pq \vert rs\rangle \{ a^\dagger_p  a^\dagger_q a_s a_r \}.
        \end{equation}
To obtain the working equation for the PT amplitudes, we substitute eqs.~\eqref{eq:h0o} and \eqref{eq:vo} into eq.~\eqref{eq:pt}.
Note that, in contrast to the PT methods with a diagonal zero-order Hamiltonian, the PT amplitudes are now obtained from a set of coupled equations.

\paragraph{Restricting the dual $\langle \tilde{\Psi} \vert$ to $\langle 0\vert$.}
If the dual state is restricted to the reference determinant of the AP1roG wavefunction, we benefit from the intermediate normalization when evaluating the overlap $\langle 0\vert \mathrm{AP1roG} \rangle$.
Analogous to PT2SDd-type methods, the second order energy can be evaluated from eq.~\eqref{eq:epta}.
Note, however, that only the doubly-excited determinants directly contribute to the energy correction.
Since the perturbation Hamiltonian is a two-electron operator, the second-order energy correction of the single excitations vanishes.
Single excitations contribute indirectly through coupling to the double excitation manifold in the PT amplitude equations.
For both including and excluding the singles projection manifold, the second-order energy $E^{(2)}$ is thus calculated from eq.~\eqref{eq:e2sdd}.
We will abbreviate the PT corrections using an off-diagonal (o) zero-order Hamiltonian and a single determinant for its dual state as PT2SDo, while the projection manifold will be indicated in parentheses (d for doubles, sd for singles and doubles, respectively).

\paragraph{Choosing $\langle \mathrm{AP1roG} \vert$ as dual $\langle \tilde{\Psi} \vert$.}
Similar to the PT methods with a diagonal $\hat{H}_0$ Hamiltonian, choosing $\langle \mathrm{AP1roG} \vert$ as dual forces us to evaluate the wavefunction overlap $\langle \mathrm{AP1roG} \vert \mathrm{AP1roG} \rangle$, which becomes prohibitive for large systems.
In order to (partially) remove the wavefunction overlap from the working equations, we follow the procedure from above and introduce a scaled zero-order Hamiltonian, where we have to modify both the diagonal ${{F}}_N^{\mathrm{d}}$ and off-diagonal ${{F}}_N^{\mathrm{o}}$ Fock operator,
        \begin{align}
            (\hat{H}_0)_N  &= \bar{{F}}_N^{\mathrm{d}} +\bar{{F}}_N^{\mathrm{o}} \nonumber \\
                           &= \sum_{p,q} \frac{f_{pq}}{\langle \mathrm{AP1roG} \vert \mathrm{AP1roG} \rangle} \{a^\dagger_p a_q\}
                            = \sum_{p,q} \bar{f}_{pq} \{a^\dagger_p a_q\}.
        \end{align}
Since we use a modified Fock operator as zero-order Hamiltonian, we have to account for it in the definition of the perturbation part,
        \begin{align}\label{eq:vno}
            \hat{V}_N^\prime  &= \hat{F}_N-\bar{{F}}_N^{\mathrm{d}}-\bar{{F}}_N^{\mathrm{o}} + \hat{W}_N^\prime \nonumber \\
                              &= \sum_{p,q} (f_{pq}-\bar{f}_{pq}) \{a^\dagger_p a_q\} + \frac{1}{2}\sum_{pqrs}{}^\prime  \langle pq \vert rs\rangle \{ a^\dagger_p  a^\dagger_q a_s a_r \}.
        \end{align}
Analogous to PT2MDd-type methods, the scaling factor in the $(\hat{H}_0)_N$ Hamiltonian is absorbed in the PT amplitudes (cf.eq.~\eqref{eq:psi1md}) so that the zero-order Hamiltonian in the PT amplitude equations eq.~\eqref{eq:pt} contains only the unscaled Fock operator, $(\hat{H}_0)_N  = \hat{{F}}_N^{\mathrm{d}} + \hat{{F}}_N^{\mathrm{o}}$, while the PT amplitudes $t_p$ are replaced by the scaled amplitudes $\bar{t}_p$.
Although we eliminated the wavefunction overlap in the second-order energy expression and in the $(\hat{H}_0)_N$ part of the PT amplitude equations, $\langle \mathrm{AP1roG} \vert \mathrm{AP1roG} \rangle$ still remains in the perturbation part (cf.~eq.~\eqref{eq:vno}).
In contrast to the PT2MDd-type methods discussed above, the perturbation $\hat{V}_N^\prime$ contains also modified off-diagonal elements in the one-electron Fock operator that do not vanish in the PT equations.
Therefore, we have to evaluate the wavefunction overlap $\langle \mathrm{AP1roG} \vert \mathrm{AP1roG} \rangle$ before we can determine the PT amplitudes.
To obtain a computationally feasible model, we will approximate the overlap $\langle \mathrm{AP1roG} \vert \mathrm{AP1roG} \rangle \approx 1 + \sum_{ia} \vert c_i^a \vert^2$, keeping only the quadratic terms in the AP1roG amplitudes.
This is usually a good approximation as the AP1roG wavefunction amplitudes are typically much smaller than 1 ($c_i^a \ll 1$).
Thus, the scaled off-diagonal Fock matrix elements can be approximated by $\bar{f}_{pq} \approx \frac{f_{pq}}{1 + \sum_{ia} \vert c_i^a \vert^2}$.
As for PT2MDd, the second-order energy $E^{(2)}$ can be determined from eq.~\eqref{eq:e2md} with $\hat{V}_N^\prime$ defined in eq.~\eqref{eq:vno}.
Note that in contrast to PT2SDo, the single excitations directly contribute to the energy correction through both the one- and two-electron operators in the perturbation Hamiltonian.
The PT models with an off-diagonal (o) $\hat{H}_0$ Hamiltonian and a multi-determinant (MD) wavefunction ($\langle \mathrm{AP1roG} \vert$) as dual will be abbreviated as PT2MDo, while the projection manifold will be again indicated in parentheses (d for doubles, sd for singles and doubles, respectively).
Note that pair-excitations are excluded in the projection manifold due to the orthogonality constraint.

\paragraph{Relation to PTb theory.}
The PT2MDo model is similar, but not equivalent, to the recently presented PTb approach.~\cite{Piotrus_PT2}
In the PTb method, the wavefunction overlap in the perturbation Hamiltonian is neglected, that is, we assume $\frac{1}{\langle \mathrm{AP1roG} \vert \mathrm{AP1roG} \rangle} \rightarrow 0$ so that $\bar{f}_{pq} \rightarrow 0$.
By neglecting the scaled components of the Fock operator, the perturbation Hamiltonian reduces to the full quantum chemical Hamiltonian $\hat{V}_N^\prime = \hat{H}_N^\prime$.
The second-order energy correction is then given as
        \begin{equation}\label{eq:eptb}
            E^{(2)} = \sum_p \bar{t}_p \langle \mathrm{AP1roG}  \vert \hat{H}_N^\prime \vert \Phi_p \rangle.
        \end{equation}
Furthermore, in PTb theory, pair excitations are not excluded in the projection manifold and the first-order correction of the wavefunction contains all double excitations.
These pair excitations do not contribute to the energy corrections eq.~\eqref{eq:eptb}.
Their contribution vanishes as the corresponding terms in eq.~\eqref{eq:eptb} equal the AP1roG amplitude equations.
However, pair excitations indirectly enter the energy correction by coupling to the remaining PT amplitudes in the PT equations as well as to the pair excitations of the AP1roG model.
In this work, we have extended the original PTb model by including also single excitations in the projection manifold.
Furthermore, we also investigate how the pair excitations in the projection manifold influence the PTb energy correction.
For that purpose we have excluded the pair excitations in the projection manifold when optimizing the PTb amplitudes.
To emphasize the order of the energy correction in PTb, these PT models are abbreviated as PT2b, while the projection manifold is indicated in parentheses (d for doubles, sd for singles and doubles, d\textbackslash p for doubles without pairs, sd\textbackslash p for singles and doubles excluding pair excitations).

\paragraph{Comment on computational scaling of the PT models.}
The computational scaling of all PT corrections discussed above is determined by the first term of eq.~\eqref{eq:pt}.
Note that we can introduce suitable intermediates for the second term in eq.~\eqref{eq:pt} so that summations are performed only once in the beginning of the calculation.
If $(\hat{H}_0)_N$ is restricted to be a diagonal one-body operator, the computational cost of the corresponding PT models scales as $\mathcal{O}(o^2v^2)$, where $o$ is the number of occupied and $v$ the number of virtual orbitals, respectively.
Choosing an off-diagonal one-body zero-order Hamiltonian $(\hat{H}_0)_N$ increases the computational cost to $\mathcal{O}(o^2v^3)$ (see also Table~\ref{tab:pt}).
Since we now have to solve a coupled set of linear equations iteratively, we have to consider an additional prefactor.
However, this prefactor is typically much smaller than $v$.
Thus, PT2SDd and PT2MDd scale similar to AP1roG (or conventional electronic structure methods like MP2), while in PT2SDo, PT2MDo, and PT2b the computational cost increases by a factor of $v$.
All PT approaches presented in this work are summarized in Table~\ref{tab:pt} for comparison.

\begin{table*}
\begin{center}
\caption{Summary of PT models with zero-order Hamiltonian $\hat{H}_0$, perturbation $\hat{V}$, dual $\langle \tilde{\Psi} \vert$, and excitation operator $\hat{T}$. All operators are defined in the text. The computational scaling is given in the last column.}\label{tab:pt}
\centering
\begin{tabular}{lccccc}
\hline\hline
Model & $\hat{H}_0$ & $\hat{V}$ & $\langle \tilde{\Psi} \vert$ & $\hat{T}$ & scaling \\ \hline
PT2SDd & $\hat{F}_N^{\mathrm{d}}$ & ${F}_N^{\mathrm{o}} + \hat{W}_N^\prime$ & $\langle 0 \vert $ & $\hat{T}_2^\prime$, $\hat{T}_1+\hat{T}_2^\prime$ & $\mathcal{O}(o^2v^2)$\\
PT2MDd & $\hat{F}_N^{\mathrm{d}}$ & ${F}_N^{\mathrm{o}} + \hat{W}_N^\prime$ & $\langle \mathrm{AP1roG} \vert $ & $\hat{T}_2^\prime$, $\hat{T}_1+\hat{T}_2^\prime$ & $\mathcal{O}(o^2v^2)$\\
PT2SDo & $\hat{F}_N^{\mathrm{d}} + \hat{F}_N^{\mathrm{o}}$ & $\hat{W}_N^\prime$ & $\langle 0 \vert $ & $\hat{T}_2^\prime$, $\hat{T}_1+\hat{T}_2^\prime$ & $\mathcal{O}(o^2v^3)$\\
PT2MDo & $\hat{F}_N^{\mathrm{d}} + \hat{F}_N^{\mathrm{o}}$ & $\hat{F}_N - \bar{F}_N + \hat{W}_N^\prime$ & $\langle \mathrm{AP1roG} \vert $& $\hat{T}_2^\prime$, $\hat{T}_1+\hat{T}_2^\prime$ & $\mathcal{O}(o^2v^3)$\\
PT2b   & $\hat{F}_N^{\mathrm{d}} + \hat{F}_N^{\mathrm{o}}$ & $\hat{H}_N^\prime$ & $\langle \mathrm{AP1roG} \vert $& $\hat{T}_2$, $\hat{T}_2^\prime$, $\hat{T}_1+\hat{T}_2$, $\hat{T}_1+\hat{T}_2^\prime$ &  $\mathcal{O}(o^2v^3)$\\
\hline\hline
\end{tabular}
\end{center}
\end{table*}

\subsection{Combining AP1roG with MP2}
In our last PT model, we will combine conventional MP2 theory with AP1roG.
Specifically, all wavefunction amplitudes corresponding to singly and doubly excited determinants with respect to $\vert 0 \rangle$ are determined from the MP2 amplitude equations.
To prevent double counting of the correlation contribution associated with electron pairs, we omit the amplitudes of all pair-excited determinants in the MP2 equations.
In MP2, the PT amplitudes are determined from a set of uncoupled equations, 
        \begin{equation}\label{eq:mp2}
            \sum_p {}^\prime t_p \langle \overline{\Phi_q} \vert (\hat{H}_0)_N \vert \Phi_p \rangle + \langle \overline{\Phi_q} \vert \hat{V}_N \vert 0 \rangle = 0,
        \end{equation}
where $\vert 0 \rangle$ is the AP1roG reference determinant (cf.~eq.~\eqref{eq:pt}).
The second-order energy corrections then accounts for correlation contributions beyond electron-pairs and can be determined from eq.~\eqref{eq:e2sdsd}, which contains both single and double excitations with respect to $\vert 0 \rangle$.
We will label this methods as AP1roG-MP2.

\subsection{Linearized Coupled Cluster Corrections}
Besides PT-type methods, dynamic correlation effects can be built in the AP1roG wavefunction \emph{a posteriori} using an exponential coupled cluster ansatz,~\cite{Kasia-lcc}
        \begin{equation}\label{eq:lcc}
            |\Psi \rangle = \exp({\hat{T}})  \vert {\rm AP1roG} \rangle,
        \end{equation}
where $\hat{T} = \sum_\nu t_\nu \hat{\tau}_\nu$ is a general cluster operator. The corresponding time-independent Schr\"odinger equation then reads
        \begin{equation}
            \hat{H} \exp(\hat{T}) \vert {\rm AP1roG} \rangle = E \exp(\hat{T}) \vert {\rm AP1roG} \rangle.
        \end{equation}
Instead of considering the full expansion of the cluster operator in the exponential function, we can include terms only linear in $\hat{T}$.
If we truncate the Baker--Campbell--Hausdorff expansion after the second term,
        \begin{equation}
            \exp(-\hat{T})\hat{H} \exp(\hat{T}) \approx \hat{H} + [\hat{H},\hat{T}],
        \end{equation}
we obtain the linearized coupled cluster Schr\"odinger equation
        \begin{equation}\label{eq:lccse}
            (\hat{H} + [\hat{H},\hat{T}]) \vert {\rm AP1roG} \rangle =  E \vert {\rm AP1roG} \rangle.
        \end{equation}
The cluster amplitudes $t_\nu$ are determined by solving a linear set of coupled equations obtained from multiplying the above equation from left by determinants of the projection manifold $\langle \nu \vert$,
        \begin{equation}
            \langle \nu \vert (\hat{H} + [\hat{H},\hat{T}]) \vert {\rm AP1roG} \rangle =  0.
        \end{equation}
The energy can be calculated by projecting against the reference determinant of $\vert {\rm AP1roG} \rangle$,
        \begin{equation}
          \langle 0 \vert (\hat{H} + [\hat{H},\hat{T}]) \vert {\rm AP1roG} \rangle = E.
        \end{equation}
Specifically, in the AP1roG-LCCD approach, the cluster operator is restricted to double excitations with respect to $\vert 0 \rangle$ as defined in eq.~\eqref{eq:t2}, while the AP1roG-LCCSD method also single excitations are included and hence $\hat{T}= \hat{T}_1+\hat{T}_2^\prime$ (cf.~eqs.\eqref{eq:t2} and \eqref{eq:t1}).
For more information concerning the choice of the cluster operator and the projection manifold, we refer the reader to ref.~\citenum{Kasia-lcc}.
Note that the computational scaling of AP1roG-LCCD/LCCSD is similar to the computational scaling of conventional LCCD/LCCSD methods, that is, $\mathcal{O}(o^2v^4)$.

\section{Computational details}\label{sec:compdetails}

\subsection{Geometry optimization and single point calculations}
All molecular structures used in the calculations of reaction energies were fully optimized in the ~\textsc{TURBOMOLE7.0} software package~\cite{turbomole-paper}, 
employing the B3LYP exchange--correlation functional~\cite{B3LYP_Becke,B3LYP} and a polarized valence triple-$\zeta$ basis set~\cite{def-TZVP,def-TZVP2}. 
The resulting geometries in xyz format are included in the Supporting Information.  

The single-point MP2, BCCD, and CCSD(T) calculations have been carried out in the~\textsc{Molpro2012.1} software package~\cite{molpro2012,molpro-WIREs}, 
while the \textsc{NWChem} software suite (in combination with the tensor contraction engine~\cite{tce_1,tce_2,tce_3,tce_4}) was used for LCCD, LCCSD, and CR-CCSD(T) calculations.
All AP1roG calculations including \textit{a posteriori} corrections of the dynamic correlation energy were performed in our in-house quantum chemistry code.
The single-point energies of all investigated reactions were evaluated using the cc-pVDZ, cc-pVTZ, and cc-pVQZ basis sets of Dunning~\cite{dunning_b} and extrapolated to the basis set limit. 
Two sets of calculations were performed, one with the canonical Hartree--Fock orbitals and another with the AP1roG optimized orbital basis.   

The dissociation curve of the C$_2$ (ground state) molecule was determined for Dunning's aug-cc-pVTZ basis set, while for N$_2$ (ground state) and BN (first excited state) the cc-pVTZ basis set was employed. 
All the dynamic energy corrections on top of AP1roG were calculated in the AP1roG optimized orbital basis. 
The points of the potential energy curves of all investigated diatomics were used for a subsequent generalized Morse function~\cite{Coxon_1992} fit to obtain the equilibrium bond lengths ($r_{\rm e}$) and potential energy depths ($D_{\rm e}$).
The harmonic vibrational frequencies ($\omega_{\rm e}$) were calculated numerically using the five-point finite difference stencil~\cite{Abramowitz}.

\subsection{Extrapolation to the basis set limit}
The basis set limit of the Hartree--Fock energy was obtained by fitting an exponential function of the form~\cite{Helgaker1997}
        \begin{equation}
            E^{\mathrm{SCF}}({X}) = E^{\mathrm{SCF}}_{\infty} + a \exp(-b X)
        \end{equation}
to the Hartree--Fock energies obtained in the cc-pVDZ, cc-pVTZ, and cc-pVQZ basis sets.
In the above equation $X$ indicates the cardinal number of the basis set (2 for D, 3 for T, etc.).
For all correlation calculations, the basis set limit of the correlation energy was obtained by a two-point fit using the fit function
        \begin{equation}
            E^{\mathrm{corr}}({X}) = E^{\mathrm{corr}}_{\infty} + a X^{-3},
        \end{equation}
as suggested in refs.~\cite{Helgaker1997,Halkier1998a}.
In the above equation, $E^{\mathrm{corr}}({X})$ indicates the correlation energy of a given method defined as $E^{\mathrm{corr}}({X})= E^{\mathrm{tot}}({X})-E^{\mathrm{SCF}}({X})$.
In the case of dynamic correction on top of AP1roG, the correlation energy is thus a sum of the correlation energy of AP1roG and of the dynamic correlation model.
Note, however, that extrapolating the sum of correlation energies or each contribution separately yields similar basis set limits of the (total) correlation energy $E^{\mathrm{corr}}_{\infty}$.
For all correlation calculations, only the correlation energies of the cc-pVTZ ($X=3$) and cc-pVQZ ($X=4$) basis sets were employed in the fitting procedure.

\section{Results and discussion}\label{sec:results}

The previously discussed dynamic correlation models are benchmarked against spectroscopic constants of multiply bonded diatomics and thermochemical data for compounds containing main-group elements.
Note that the former test systems are dominated by static/nondynamic electron correlation effects, while dynamic electron correlation becomes important for the latter.
Both test sets allow us to assess the accuracy of the proposed dynamic corrections for strongly- and weakly correlated molecular systems.

\subsection{Molecular systems dominated by static/nondynamic correlation}\label{sec:results1}
We will first briefly discuss the performance and trends of the proposed PT corrections in modeling dissociation processes of three diatomic molecules, C$_2$, N$_2$, and BN.
Their electronic structures have been extensively discussed in the literature and we refer the interested reader to, for instances, refs.~\cite{Peterson1993-homo,Peterson1995,Sherill_C2,entanglement_bonding_2013,Dunning_C2} for more details.
Table~\ref{tab:diatomics} summarizes the spectroscopic constants for the C$_2$, N$_2$, and BN molecules.
The corresponding potential energy surfaces are shown in the Supplementary Information.
For C$_2$ and N$_2$, MRCI-SD data by Peterson \textit{et al.}~\cite{Peterson1993-homo} is taken as reference, while MRCI-SD+Q results are used as reference for the BN molecule~\cite{Peterson1995}.

\begin{table}
\begin{center}
    \caption{Spectroscopic constants for the dissociation of the C$_2$, N$_2$, and BN molecule for different quantum chemistry methods. The differences are with respect to MRCI-SD reference data~\cite{Peterson1993-homo} for C$_2$ and N$_2$, and with respect to MRCI-SD+Q reference data~\cite{Peterson1995} for BN. $E_e$: ground state energy at $r_e$.}\label{tab:diatomics}
{\scriptsize
    \begin{tabular}{l|l|ccr@{(}lr@{(}l} \hline\hline
        & Method & \multicolumn{1}{c}{$E_e$ [$E_{h}$]} & \multicolumn{1}{c}{$r_e$ [\AA{}]} & \multicolumn{2}{c}{$D_e$ [$\frac{\rm kcal}{\rm mol}$]} & \multicolumn{2}{c}{$\omega_e$ [cm$^{-1}$]}    \\ \hline \hline
\multirow{17}{*}{C$_2$} & AP1roG & $-$75.58569 & 1.227($-$0.025) & 132.9&$-$8.2)      & 1780&$-$56) \\
             &  AP1roG-PT2SDd(d) & $-$75.80222 & 1.251($+$0.001) & 160.4&$+$19.3)     & 1889&$+$53) \\
             &  AP1roG-PT2SDd(sd)& $-$75.81041 & 1.249($-$0.003) & 154.6&$+$13.5)     & 1915&$+$79) \\
             &  AP1roG-PT2MDd(d) & $-$75.77832 & 1.239($-$0.013) & 121.6&$-$19.5)$^*$ & 1940&$+$104)\\
             &  AP1roG-PT2MDd(sd)& $-$75.78630 & 1.238($-$0.014) & 115.6&$-$25.5)$^*$ & 1963&$+$127)\\
             &  AP1roG-PT2SDo(d) & $-$75.81778 & 1.242($-$0.010) & 156.5&$+$15.4)     & 1919&$+$83) \\
             &  AP1roG-PT2SDo(sd)& $-$75.81783 & 1.242($-$0.010) & 156.4&$+$15.3)     & 1919&$+$83) \\
             &  AP1roG-PT2MDo(d) & $-$75.79032 & 1.231($-$0.021) & 125.9&$-$15.2)$^*$ & 2010&$+$174)\\
             &  AP1roG-PT2MDo(sd)& $-$75.79108 & 1.230($-$0.022) & 119.9&$-$21.2)$^*$ & 2019&$+$183)\\
             &  AP1roG-PT2b(d)   & $-$75.78350 & 1.235($-$0.017) & 127.2&$-$13.9)$^*$ & 1938&$+$102)\\
             &  AP1roG-PT2b(sd)  & $-$75.79400 & 1.228($-$0.024) & 113.0&$-$28.1)$^*$ & 2049&$+$213)\\
&  AP1roG-PT2b(d\textbackslash p)& $-$75.78381 & 1.231($-$0.021) & 123.9&$-$17.2)$^*$ & 2016&$+$180)\\
& AP1roG-PT2b(sd\textbackslash p)& $-$75.79370 & 1.228($-$0.024) & 112.9&$-$28.2)$^*$ & 2048&$+$212)\\
& AP1roG-LCCD~\cite{Kasia-lcc}   & $-$75.81125 & 1.240($-$0.012) & 139.3&$-$1.8)      & 1916&$+$80) \\
&  AP1roG-LCCSD~\cite{Kasia-lcc} & $-$75.81257 & 1.240($-$0.012) & 143.0&$+$1.9)      & 1926&$+$90) \\
             &  NEVPT2~\cite{pawel_jpca_2014}    & $-$75.78829 & 1.244($-$0.008)      & 148.0&$+$6.9)  & 1886&$+$50) \\
             &  CR-CCSD(T)       & $-$75.80484 & 1.242($-$0.010) & 152.1&$+$9.0)      & 1989&$+$153) \\ \cline{2-8}
&MRCI-SD    \cite{Peterson1993-homo} & $-$75.78079 & 1.252 & \multicolumn{2}{c}{141.1} & \multicolumn{2}{c}{1836}\\ \hline \hline

\multirow{18}{*}{N$_2$} & AP1roG &  $-$109.12686 & 1.087($+$0.083) & 255.9&$-$38.0)     & 2435&$+$94) \\
             &  AP1roG-PT2SDd(d) &  $-$109.37326 & 1.094($-$0.010) & 249.8&$-$31.9)     & 2301&$-$40) \\
             &  AP1roG-PT2SDd(sd)&  $-$109.37383 & 1.093($-$0.011) & 233.4&$+$15.5)     & 2299&$-$42) \\
             &  AP1roG-PT2MDd(d) &  $-$109.35918 & 1.104($+$0.000) & 210.3&$-$7.6)$^*$  & 2744&$+$403)\\
             &  AP1roG-PT2MDd(sd)&  $-$109.35988 & 1.105($+$0.001) & 202.0&$-$15.1)$^*$ & 2779&$+$438)\\
             &  AP1roG-PT2SDo(d) &  $-$109.39838 & 1.110($+$0.006) & 246.9&$+$29.0)     & 2278&$-$63) \\
             &  AP1roG-PT2SDo(sd)&  $-$109.39838 & 1.108($+$0.004) & 248.8&$+$30.9)     & 2254&$-$87) \\
             &  AP1roG-PT2MDo(d) &  $-$109.38106 & 1.089($-$0.015) & 220.8&$+$2.9)$^*$  & 2200&$-$141)\\
             &  AP1roG-PT2MDo(sd)&  $-$109.38132 & 1.090($-$0.014) & 216.7&$-$1.0)$^*$  & 2203&$-$138)\\
             &  AP1roG-PT2b(d)   &  $-$109.37490 & 1.092($-$0.012) & 219.2&$+$1.3)$^*$  & 2246&$-$95) \\
             &  AP1roG-PT2b(sd)  &  $-$109.37582 & 1.095($-$0.009) & 207.1&$-$10.8)$^*$ & 2272&$-$69) \\
&  AP1roG-PT2b(d\textbackslash p)&  $-$109.37431 & 1.092($-$0.012) & 218.9&$+$1.0)$^*$  & 2240&$-$101)\\
& AP1roG-PT2b(sd\textbackslash p)&  $-$109.37525 & 1.094($-$0.010) & 206.8&$-$11.1)$^*$ & 2263&$-$78) \\
             & AP1roG-fLCCD      &  $-$109.39671 & 1.102($-$0.002) & 206.7&$-$11.2)     & 2334&$-$7)  \\
             & AP1roG-fLCCSD     &  $-$109.39871 & 1.103($-$0.001) & 211.6&$-$6.3)      & 2337&$-$4)  \\
             & CASSCF~\cite{Peterson1993-homo}            &  $-$109.13190 & 1.106($+$0.002) & 211.6&$-$6.3)      & 2340&$+$1)  \\ 
             & CR-CCSD(T)        &  $-$109.39632 & 1.101($-$0.003) & 229.7&$+$11.8)$^*$ & 2397&$+$56) \\ \cline{2-8}
&MRCI-SD \cite{Peterson1993-homo}&  $-$109.36162 & 1.104 & \multicolumn{2}{c}{217.9} & \multicolumn{2}{c}{2341} \\ \hline \hline

\multirow{19}{*}{BN} & AP1roG    &  $-$79.07582  & 1.304($+$0.019) & 114.3&$-$40.1)      & 1630&$-$52) \\
             &  AP1roG-PT2SDd(d) &  $-$79.28333  & 1.283($-$0.002) & 181.3&$+$26.9)      & 1751&$+$68) \\
             &  AP1roG-PT2SDd(sd)&  $-$79.29473  & 1.271($-$0.014) & 166.1&$+$11.7)$^*$  & 1784&$+$102)\\
             &  AP1roG-PT2MDd(d) &  $-$79.26001  & 1.273($-$0.012) & 160.7&$+$6.3)       & 1778&$+$96) \\
             &  AP1roG-PT2MDd(sd)&  $-$79.27571  & 1.264($-$0.021) & 144.5&$-$9.9)       & 1760&$+$78) \\
             &  AP1roG-PT2SDo(d) &  $-$79.29432  & 1.282($-$0.003) & 169.9&$+$15.5)      & 1718&$+$36) \\
             &  AP1roG-PT2SDo(sd)&  $-$79.29427  & 1.281($-$0.004) & 169.7&$+$15.3)$^*$  & 1719&$+$37) \\
             &  AP1roG-PT2MDo(d) &  $-$79.26814  & 1.274($-$0.011) & 158.5&$+$4.1)$^*$   & 1743&$+$61) \\
             &  AP1roG-PT2MDo(sd)&  $-$79.27584  & 1.269($-$0.016) & 139.0&$-$15.4)$^*$  & 1725&$+$43) \\
             &  AP1roG-PT2b(d)   &  $-$79.25960  & 1.273($-$0.012) & 152.6&$-$1.8)$^*$   & 1739&$+$57) \\
             &  AP1roG-PT2b(sd)  &  $-$79.27961  & 1.260($+$0.025) & 141.9&$-$12.5)$^*$  & 1746&$+$64) \\
&  AP1roG-PT2b(d\textbackslash p)&  $-$79.25938  & 1.273($+$0.012) & 156.6&$+$2.2)$^*$   & 1741&$+$59) \\
& AP1roG-PT2b(sd\textbackslash p)&  $-$79.27934  & 1.261($-$0.024) & 142.3&$-$12.1)$^*$  & 1746&$+$64) \\
&  AP1roG-LCCD~\cite{Kasia-lcc}  &  $-$79.28205  & 1.277($-$0.008) & 174.1&$-$19.8)      & 1734&$+$52) \\
&  AP1roG-LCCSD~\cite{Kasia-lcc} &  $-$79.28975  & 1.275($-$0.010) & 178.6&$+$24.2)      & 1749&$+$67) \\
             &        CCSD(T)~\cite{Paldus2006}    &  ---          & 1.244($-$0.008) & \multicolumn{2}{c}{---} & 1886&$+$50) \\
             & CR-CCSD(T)        &  $-$79.28080  & 1.275($-$0.010) & 164.9&$+$10.5) & 1726&$+$44) \\
             & RMR CCSD(T) \cite{Paldus2006} & --- & 1.284($-$0.001) & \multicolumn{2}{c}{---} & 1681&$+$1)\\\cline{2-8}
 & MRCI-SD+Q \cite{Peterson1995} & $-$79.26794   & 1.285 & \multicolumn{2}{c}{154.4} & \multicolumn{2}{c}{1682}\\ \hline \hline
    \end{tabular} 
}
\end{center}
\begin{tablenotes}\footnotesize
\item $^*$ Estimated dissociation energy.\\
\end{tablenotes}
\end{table}
In general, PT2SDd and PT2SDo provide equilibrium bond lengths and vibrational frequencies that agree well with the corresponding reference values for all investigated diatomics.
Furthermore, addition of single excitations in the excitation operator $\hat{T}$ does not increase the accuracy of PT2SDd(d) or PT2SDo(d).
In contrast to PT2SDd and PT2SDo, the performance of PT models using a multi-determinant dual state (PT2MDd, PT2MDo, and PT2b) is less satisfying.
Specifically, these methods provide equilibrium bond distances and vibrational frequencies that deviate most from MRCI-SD/MRCI-SD+Q reference data.
Similar to PT2SDd and PT2SDo, addition of single excitations worsens spectroscopic constants and hence the excitation operator should be restricted to double excitations only.
Furthermore, including pair excitations in $\hat{T}$ (PT2b-type methods) does not significantly affect spectroscopic constants and both models (with and without pair excitations) yield similar values for $r_e$ and $\omega_e$.

In contrast to equilibrium bond distances and vibrational frequencies, the accurate prediction of dissociation energies $D_e$ is more challenging.
In general, none of the proposed PT models can reliably predict potential energy well depths, which differ by approximately 15 to 30 kcal/mol from MRCI-SD/MRCI-SD+Q reference data.
Note that most PT corrections (PT2MD and PT2b) diverge in the vicinity of dissociation and hence only an estimated dissociation energy is given in Table~\ref{tab:diatomics} (indicated by the $*$ in the Table).

The performance of the LCCD/LCCSD correction on top of AP1roG is more robust in predicting spectroscopic constants for all investigated diatomics.
Although AP1roG-LCCSD yields $r_e$ and $\omega_e$ that deviate more from MRCI-SD/MRCI-SD+Q reference values compared to PT2SDd and PT2SDo results, an LCC correction allows us to reliably model dissociation energies for C$_2$ and N$_2$, while the errors in $D_e$ increase for BN (differences are 2 kcal/mol for C$_2$, 6 kcal/mol for N$_2$, and 24 kcal/mol for BN).
Note that for the N$_2$ molecule the coupled cluster doubles amplitudes for excitations within the valence shell (vs) have been kept frozen in the vicinity of dissociation ($r>2.0$ \AA{} with $t_{r>2.0}^{\textrm{vs}}=t_{r=2.0}^{\rm vs}$) to prevent divergencies in the dissociation limit and to obtain a smooth potential energy surface.
This is indicate by fLCCD/fLCCSD in Table~\ref{tab:diatomics}.
The corresponding AP1roG-fLCCSD spectroscopic constants agree well with CASSCF results by Peterson \textit{et al.}~\cite{Peterson1993-homo} and are in general closest to MRCI-SD reference values.

\subsection{Molecular systems dominated by dynamic correlation}\label{sec:results2}

Our second test case includes 15 reactions containing main-group elements summarized in Table~\ref{tab:reactions}.
For all studied systems, the reaction energies obtained from CR-CCSD(T) calculations are taken as reference values.
Furthermore, we will focus on two different molecular orbital basis sets used in our AP1roG calculations and in all \textit{a posteriori} corrections for dynamic correlation with an AP1roG reference function; these are optimized natural AP1roG orbitals and canonical Hartree--Fock orbitals. 

\newcounter{foo}
\begin{table} 
\begin{center}
\caption{List of reactions containing closed-shell molecules.}\label{tab:reactions}
\centering
\begin{tabular}{lc}
\hline\hline
\ce{F2 + H2 -> 2HF}         & \{\refstepcounter{foo}\thefoo\}\label{r1}\\
\ce{F2O + H2 -> F2 + H2O}     & \{\refstepcounter{foo}\thefoo\}\label{r2}\\
\ce{H2O2 + H2 -> 2H2O}     & \{\refstepcounter{foo}\thefoo\}\label{r3}\\
\ce{N2 + 3H2 -> 2NH3}     & \{\refstepcounter{foo}\thefoo\}\label{r4}\\
\ce{N2O + H2 -> N2 + H2O}   & \{\refstepcounter{foo}\thefoo\}\label{r5}\\
\ce{C2H2 + H2 -> C2H4}      & \{\refstepcounter{foo}\thefoo\}\label{r6}\\
\ce{BH3 + 3HF -> BF3 + 3 H2} & \{\refstepcounter{foo}\thefoo\}\label{r7}\\
\ce{CO + H2O -> CO2 + H2}   & \{\refstepcounter{foo}\thefoo\}\label{r8}\\
\ce{CO + 3H2 -> CH4 + H2O}  & \{\refstepcounter{foo}\thefoo\}\label{r9}\\
\ce{2BH3 -> B2H6}          & \{\refstepcounter{foo}\thefoo\}\label{r10}\\
\ce{2H2O -> (H2O)2}        & \{\refstepcounter{foo}\thefoo\}\label{r11}\\
\ce{2HF -> (HF)2}          & \{\refstepcounter{foo}\thefoo\}\label{r12}\\
\ce{HCOOH -> CO2 + H2}      & \{\refstepcounter{foo}\thefoo\}\label{r13}\\
\ce{CO + CH4 -> CH3CHO}     & \{\refstepcounter{foo}\thefoo\}\label{r14}\\
\ce{2NH3 -> (NH3)2}        & \{\refstepcounter{foo}\thefoo\}\label{r15}\\
\hline\hline
\end{tabular}
\label{tbl:reactions}
\end{center}
\end{table}

\begin{figure*}[t]
\includegraphics[width=0.9\textwidth]{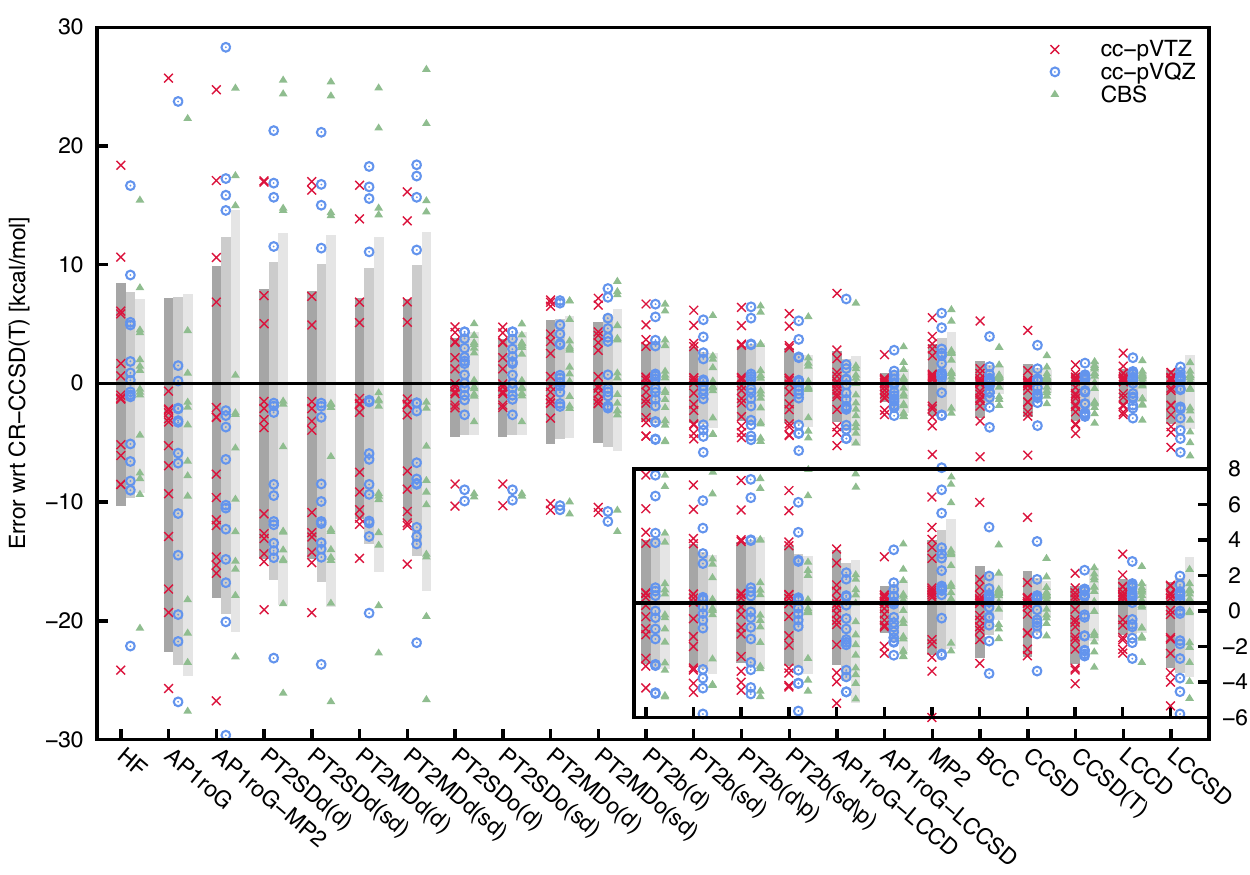}
\caption{Errors in the optimized AP1roG orbital basis.}\label{fig:err1}
\end{figure*}

%
%
\renewcommand{\arraystretch}{1.2}
\begin{table*} 
{\tiny
\caption{Reaction energies of~\{\ref{r1}\}--\{\ref{r15}\} obtained from different theoretical models using a cc-pVQZ basis set (in kcal/mol).
All dynamic energy corrections on top of AP1roG are determined for the optimized natural AP1roG basis.
Only the CR-CCSD(T) reference energy and the energy difference with respect to CR-CCSD(T) $(E^{\mathrm{CR-CCSD(T)}}_p-E_p^{\mathrm{method}})$ is given in the Table.
ME: mean error ($\mathrm{ME}=\sum_p^N (E^{\mathrm{CR-CCSD(T)}}_p-E_p^{\mathrm{method}})/N$);
RMSE: root mean square error ($\mathrm{RMSE}=\sqrt{\sum_p^N (E^{\mathrm{CR-CCSD(T)}}_p-E_p^{\mathrm{method}})^2/N}$);
MAE: mean absolute error ($\mathrm{MAE}=\sum_p^N \vert E^{\mathrm{CR-CCSD(T)}}_p-E_p^{\mathrm{method}} \vert /N$);
max AE: maximum absolute error ($\mathrm{max\,AE}=\max{(\{\vert E^{\mathrm{CR-CCSD(T)}}_p-E_p^{\mathrm{method}}\vert }\}$).
}\label{tab:err1}
\rowcolors{2}{black!5}{white}
\begin{tabular}{l d{1}d{1}d{1}d{1}d{1}d{1}d{1}d{1}d{1}d{1}d{1}d{1}d{1}d{1}d{1}||d{1}d{1}d{1}d{1}}
\hline
\hline
\diaghead{AP1roG-LCCSDLL}{\tiny Method}{\tiny Reaction\\\quad}
 & \multicolumn{1}{c}{\{\ref{r1}\}} & \multicolumn{1}{c}{\{\ref{r2}\}} &\multicolumn{1}{c}{\{\ref{r3}\}} &\multicolumn{1}{c}{\{\ref{r4}\}} &\multicolumn{1}{c}{\{\ref{r5}\}} &\multicolumn{1}{c}{\{\ref{r6}\}} &\multicolumn{1}{c}{\{\ref{r7}\}} &\multicolumn{1}{c}{\{\ref{r8}\}} &\multicolumn{1}{c}{\{\ref{r9}\}} &\multicolumn{1}{c}{\{\ref{r10}\}} &\multicolumn{1}{c}{\{\ref{r11}\}} &\multicolumn{1}{c}{\{\ref{r12}\}} &\multicolumn{1}{c}{\{\ref{r13}\}} & \multicolumn{1}{c}{\{\ref{r14}\}} & \multicolumn{1}{c||}{\{\ref{r15}\}} & \multicolumn{1}{c}{ME} & \multicolumn{1}{c}{RMSE} & \multicolumn{1}{c}{MAE} & \multicolumn{1}{c}{max AE}\\ \hline
CR-CCSD(T)  & -135.8 & -70.9 & -88.1 & -40.5 & -83.2 & -50.0  &  -92.5  &  -6.0   &  -66.5 & -44.0  &  -4.7   &  -4.4  &  2.4   &  3.8    &  -2.7  &  -     &   -    &   -    &  -   \\ \hline
HF          &   9.1  &  5.2  & 4.9   & -6.6  & 16.7  & 0.9    &  0.3    &  -5.1   &  -9.0  & -22.1  &  -1.0   &  -0.7  &  1.9   &  -8.2   &  -1.1  &  -1.0  &  8.7   &  6.2   &  22.1\\
AP1roG      & -19.5  & -5.9  & -14.5 & -26.8 & 23.7  & 0.2    &  -3.1   &  -10.9  &  -21.7 & -31.9  &  1.5    &  -2.1  &  -3.3  &  -6.7   &  -2.2  &  -8.2  &  15.5  &  11.6  &  31.9\\
AP1roG-MP2  & 15.9   & 14.6  & 17.3  & -10.5 & 28.3  & -12.3  &  -20.1  &  -29.6  &  -3.7  & -16.8  &  -10.3  &  -2.7  &  -6.4  &  -14.8  &  -2.3  &  -3.6  &  15.9  &  13.7  &  29.6\\
PT2SDd(d)   & 15.7   & 11.5  & 16.9  & -14.7 & 21.3  & -11.9  &  -8.5   &  -23.1  &  -9.4  & -14.1  &  -11.6  &  -1.9  &  -2.4  &  -13.5  &  -1.7  &  -3.2  &  13.4  &  11.9  &  23.1\\
PT2SDd(sd)  & 15.0   & 11.4  & 16.8  & -14.6 & 21.2  & -11.8  &  -8.5   &  -23.7  &  -9.9  & -13.9  &  -11.6  &  -1.9  &  -2.8  &  -13.4  &  -1.7  &  -3.3  &  13.4  &  11.9  &  23.7\\
PT2MDd(d)   & 15.6   & 11.1  & 16.5  & -9.4  & 18.3  & -8.5   &  -6.4   &  -19.4  &  -5.9  & -12.9  &  -11.6  &  -1.5  &  -1.5  &  -11.8  &  -1.4  &  -1.9  &  11.6  &  10.1  &  19.4\\
PT2MDd(sd)  & 15.7   & 11.2  & 17.5  & -8.5  & 18.4  & -8.4   &  -6.7   &  -21.8  &  -8.0  & -13.5  &  -12.1  &  -1.7  &  -2.3  &  -12.9  &  -1.6  &  -2.3  &  12.3  &  10.7  &  21.8\\
PT2SDo(d)   & 3.0    & 4.3   & 0.7   & 1.9   & 3.8   & 0.0    &  -9.9   &  -1.4   &  4.3   & -9.0   &  1.7    &  -1.0  &  -2.7  &  2.2    &  -0.4  &  -0.2  &  4.2   &  3.1   &  9.9 \\
PT2SDo(sd)  & 2.9    & 4.3   & 0.7   & 1.8   & 3.7   & 0.0    &  -9.8   &  -1.4   &  4.3   & -9.0   &  1.7    &  -1.0  &  -2.6  &  2.3    &  -0.4  &  -0.2  &  4.2   &  3.1   &  9.8 \\
PT2MDo(d)   & 7.0    & 6.7   & 3.4   & 4.2   & 3.3   & 0.7    &  -10.3  &  -0.2   &  4.9   & -10.6  &  0.1    &  -1.9  &  -0.8  &  0.4    &  -1.7  &  0.3   &  5.1   &  3.7   &  10.6\\
PT2MDo(sd)  & 8.0    & 7.2   & 4.6   & 5.5   & 3.6   & 0.7    &  -10.8  &  -1.9   &  3.9   & -11.6  &  -0.4   &  -2.1  &  -1.1  &  -0.7   &  -1.9  &  0.2   &  5.5   &  4.3   &  11.6\\
PT2b(d)     & -1.4   & -0.8  & -1.9  & -4.7  & -4.7  & -0.4   &  5.6    &  6.7    &  -3.3  & -3.2   &  0.8    &  0.4   &  3.7   &  3.1    &  0.6   &  0.0   &  3.4   &  2.7   &  6.7 \\
PT2b(sd)    & -1.3   & -0.6  & -0.9  & -3.5  & -4.4  & -0.2   &  5.4    &  3.9    &  -5.8  & -3.9   &  0.2    &  0.2   &  2.6   &  2.1    &  0.3   &  -0.4  &  3.0   &  2.4   &  5.8 \\
PT2b(d\textbackslash p)   
            & -1.3   & -0.8  & -1.9  & -4.8  & -4.4  & -0.5   &  5.5    &  6.5    &  -3.1  & -3.5   &  0.8    &  0.4   &  3.3   &  3.2    &  0.6   &  0.0   &  3.3   &  2.7   &  6.5 \\
PT2b(sd\textbackslash p)  
            & -1.2   & -0.6  & -0.9  & -3.5  & -4.2  & -0.3   &  5.3    &  3.7    &  -5.6  & -4.2   &  0.2    &  0.1   &  2.2   &  2.2    &  0.3   &  -0.4  &  3.0   &  2.3   &  5.6 \\
AP1roG-LCCD & -2.1   & 1.6   & -0.8  & -4.6  & 7.1   & -2.2   &  -1.9   &  -4.6   &  -1.3  & -3.9   &  0.4    &  -0.1  &  -3.5  &  1.2    &  0.3   &  -1.0  &  3.1   &  2.4   &  7.1 \\
AP1roG-LCCSD& -1.8   & -1.3  & -0.8  & -2.7  & -2.0  & -1.0   &  -0.1   &  1.0    &  0.4   & -1.5   &  0.7    &  0.1   &  -2.0  &  2.8    &  0.4   &  -0.5  &  1.5   &  1.2   &  2.8 \\
MP2         & 5.9    & 0.9   & 2.6   & -2.7  & -7.1  & -2.7   &  2.4    &  4.7    &  -0.8  & 0.6    &  0.7    &  0.4   &  2.9   &  1.7    &  0.8   &  0.7   &  3.1   &  2.5   &  7.1 \\
BCCD        & 1.4    & 0.8   & 0.4   & -0.7  & 4.0   & 0.4    &  0.0    &  -2.0   &  -0.8  & -3.7   &  0.3    &  0.2   &  -1.2  &  -0.2   &  0.4   &  0.0   &  1.6   &  1.1   &  4.0 \\
CCSD        &  1.2   &  0.2  &  0.3  & -1.0  &  3.2  &   0.2  &   0.4   &  -1.6   &  -1.2  &  -3.6  &  0.3    &  0.2   &  -0.9  &  -0.4   &  0.4   &  -0.1  &  1.4   &  1.0   &  3.6 \\
CCSD(T)     & -1.7   & -2.5  & -1.7  & -2.7  & -2.5  & -0.6   &  1.7    &  0.7    &  -2.8  & -0.8   &  0.6    &  0.4   &  0.4   &  0.3    &  0.7   &  -0.7  &  1.6   &  1.3   &  2.8 \\
LCCD        & -2.9   & 0.3   & -1.1  & 0.0   & 2.2   & 0.8    &  -0.8   &  -2.1   &  1.0   & -1.2   &  0.4    &  0.3   &  -1.8  &  0.9    &  0.5   &  -0.2  &  1.4   &  1.1   &  2.9 \\
LCCSD       & -5.8   & -4.6  & -2.9  & -2.1  & -3.9  & 0.2    &  1.4    &  0.5    &  -2.0  & -0.4   &  0.5    &  0.4   &  -0.6  &  0.9    &  0.6   &  -1.2  &  2.5   &  1.8   &  5.8 \\
\hline
\hline
\end{tabular}
}
\end{table*}
\subsubsection{Optimized natural AP1roG orbitals}
Figure~\ref{fig:err1} shows the reaction energies and root mean square errors (RMSEs) with respect to CR-CCSD(T) reference data (grey bars) for all investigated methods in the optimized AP1roG basis (see Table~\ref{tab:err1} for the definition of all error measures).
Compared to Hartree--Fock data, AP1roG predicts reaction energies that differ more from the CR-CCSD(T) reference.
In general, AP1roG underestimates reaction energies and yields an RMSE of approximately 15 kcal/mol (see also Table~\ref{tab:err1} for cc-pVQZ).
This discrepancy can be attributed to electron correlation effects beyond electron pairs that are not accounted for in the AP1roG method.
Inclusion of open-shell determinants in the electronic wavefunction improves reaction energies.
However, the performance of the PT models strongly depends on the choice of the zero-order Hamilton and the perturbation.
Specifically, all PT methods with a diagonal one-electron Hamiltonian (PT2SDd and PT2MDd) provide slightly improved reaction energies with an RME of around 13 kcal/mol (see also Table~\ref{tab:err1} for cc-pVQZ).
Yet, these methods are strongly basis-set dependent if optimized AP1roG orbitals are used as orbital basis: the RMSE gradually increases with the basis set size (see Figure~\ref{fig:err1}).
An \textit{a posteriori} MP2 correction in the AP1roG basis predicts the largest RMSE of approximately 16 kcal/mol and therefore does not allow us to accurately model reaction energies.
Choosing an off-diagonal one-electron zero-order Hamiltonian (PT2SDo and PT2MDo) in the PT model further reduces the error with respect to the CCSD(T) reference with an RMSE of approximately 4-5 kcal/mol.
The error in reaction energies predicted by PT corrections can be minimized if the perturbation Hamiltonian of PT2MDo (see Table~\ref{tab:pt}) is replaced by the full quantum chemical Hamiltonian $\hat{H}_N$ as in all PT2b-type methods ($\mathrm{RMSE} \approx 3.0$ kcal/mol), where the accuracy is similar to MP2 theory (see Table~\ref{tab:err1}).
Note that inclusion of single excitations in the excitation operator ($\hat{T}=\hat{T}_1+\hat{T}_2^\prime$) does not affect the accuracy of the PT corrections in predicting reaction energies (see also Table~\ref{tab:err1}). Only the performance of PT2b is slightly improved.

For all investigated basis sets, the best agreement with CR-CCSD(T) reference data is obtained for the linearized coupled cluster correction including singles and doubles (see Figure~\ref{fig:err1}) with an RMSE of 1.5 kcal/mol and a mean absolute error of 1.2 kcal/mol.
Note that the errors of AP1roG-LCCSD is similar to those of conventional CC methods like BCC, CCSD, CCSD(T), and LCCD.
Only LCCSD performs worse with an RMSE of 2.5 kcal/mol and a mean absolute error of 1.8 kcal/mol.
We should emphasize that, unlike the LCC method with an Hartree--Fock reference function, single excitations in the LCC corrections on top of AP1roG are particularly important and allow us to approach chemical accuracy (approximately 1 kcal/mol) when modeling thermochemistry of main-group compounds.

\begin{figure*}[t]
\includegraphics[width=0.9\textwidth]{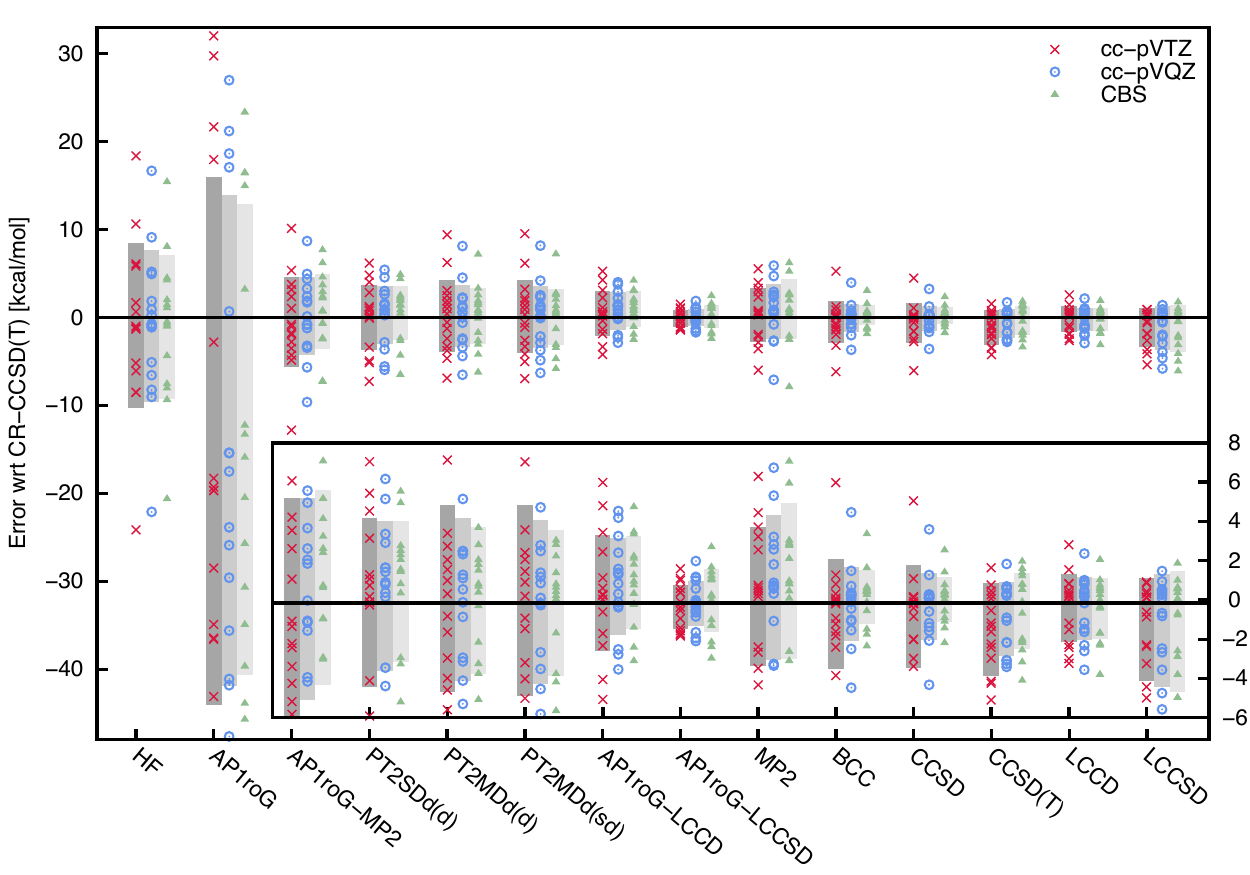}
\caption{Errors in the canonical Hartree--Fock basis.}\label{fig:err2}
\end{figure*}

\subsubsection{Canonical Hartree--Fock orbitals}
If canonical Hartree--Fock orbitals are chosen as orbital basis, we can use the diagonal structure of the Fock matrix and simplify the PT amplitude equations.
Specifically for all PT models, the PT amplitude equations are transformed into a set of uncoupled linear equations as $\hat{F}_N^{\mathrm{o}}=0$ (cf.~Table~\ref{tab:pt}).
Furthermore, single excitations do not contribute (either directly in the energy expression or indirectly through coupling to the doubles manifold) if the AP1roG reference determinant is taken as dual $\langle \tilde{\Psi} \vert$.
Thus, the PT2SDd and PT2SDo methods become equivalent (with and without single excitations) and only the PT2SDd(d) results are shown in Table~\ref{tab:err2} and Figure~\ref{fig:err2}.
Similarly, the PT2MDd and PT2MDo models are equivalent to each other.
In contrast to PT2SD-type approaches, single excitations do contribute directly to the second-order energy correction as their contribution is determined from eq.~\eqref{eq:e2md}, which also contains one-electron terms.
Finally, in PT2b-type methods, the coupling to pair-excited determinants vanishes in the PT amplitude equations as they are coupled through off-diagonal Fock matrix elements.
Furthermore, the perturbation Hamiltonian $\hat{H}_N^\prime$ of PT2b reduces to $\hat{V}_N^\prime$ and both the amplitude equations and second-order energy expression are equivalent to those in PT2MD-type methods.
If $\langle \mathrm{AP1roG} \vert$ is chosen as dual, all PT methods coincide in the canonical Hartree--Fock basis.
Therefore, Table~\ref{tab:err2} and Figure~\ref{fig:err2} show only the PT2MDd(d) and PT2MDd(sd) results.
The canonical Hartree--Fock basis thus allows us to directly assess how the choice of the dual state affects the performance of the PT models.

%
%
\renewcommand{\arraystretch}{1.2}
\begin{table*} 
\begin{center}
{\tiny
\caption{Reaction energies of~\{\ref{r1}\}--\{\ref{r15}\} obtained from different theoretical models using a cc-pVQZ basis set (in kcal/mol). All dynamic energy corrections on top of AP1roG are determined for canonical Hartree--Fock orbitals.
Only the CR-CCSD(T) reference energy and the energy difference with respect to CR-CCSD(T) $(E^{\mathrm{CR-CCSD(T)}}_p-E_p^{\mathrm{method}})$ is given in the Table.
ME: mean error ($\mathrm{ME}=\sum_p^N (E^{\mathrm{CR-CCSD(T)}}_p-E_p^{\mathrm{method}})/N$);
RMSE: root mean square error ($\mathrm{RMSE}=\sqrt{\sum_p^N (E^{\mathrm{CR-CCSD(T)}}_p-E_p^{\mathrm{method}})^2/N}$);
MAE: mean absolute error ($\mathrm{MAE}=\sum_p^N \vert E^{\mathrm{CR-CCSD(T)}}_p-E_p^{\mathrm{method}} \vert /N$);
max AE: maximum absolute error ($\mathrm{max\,AE}=\max{(\{\vert E^{\mathrm{CR-CCSD(T)}}_p-E_p^{\mathrm{method}}\vert }\}$).
}\label{tab:err2}
\rowcolors{2}{black!5}{white}
\begin{tabular}{l d{1}d{1}d{1}d{1}d{1}d{1}d{1}d{1}d{1}d{1}d{1}d{1}d{1}d{1}d{1}||d{1}d{1}d{1}d{1}}
\hline
\hline
\diaghead{AP1roG-LCCSDLL}{\tiny Method}{\tiny Reaction\\\quad}
& \multicolumn{1}{c}{\{\ref{r1}\}} & \multicolumn{1}{c}{\{\ref{r2}\}} &\multicolumn{1}{c}{\{\ref{r3}\}} &\multicolumn{1}{c}{\{\ref{r4}\}} &\multicolumn{1}{c}{\{\ref{r5}\}} &\multicolumn{1}{c}{\{\ref{r6}\}} &\multicolumn{1}{c}{\{\ref{r7}\}} &\multicolumn{1}{c}{\{\ref{r8}\}} &\multicolumn{1}{c}{\{\ref{r9}\}} &\multicolumn{1}{c}{\{\ref{r10}\}} &\multicolumn{1}{c}{\{\ref{r11}\}} &\multicolumn{1}{c}{\{\ref{r12}\}} &\multicolumn{1}{c}{\{\ref{r13}\}} & \multicolumn{1}{c}{\{\ref{r14}\}} & \multicolumn{1}{c||}{\{\ref{r15}\}} & \multicolumn{1}{c}{ME} & \multicolumn{1}{c}{RMSE} & \multicolumn{1}{c}{MAE} & \multicolumn{1}{c}{max AE}\\ \hline
CR-CCSD(T)  & -135.8 & -70.9 & -88.1 & -40.5 & -83.2 & -50.0  &  -92.5  &  -6.0   &  -66.5 & -44.0  &  -4.7   &  -4.4  &  2.4   &  3.8    &  -2.7  &  -     &   -    &   -    &  -   \\ \hline
HF           &  9.1  &   5.2  &   4.9  &  -6.6  &  16.7  &   0.9  &   0.3  &  -5.1  &  -9.0  & -22.1  &  -1.0  &  -0.7  &   1.9  &  -8.2  &  -1.1  &  -1.0  &   8.7  &   6.2  &  22.1 \\
AP1roG       &  0.7  &  27.0  &  17.1  & -47.6  &  21.2  & -15.4  & -41.8  & -15.4  & -41.1  & -35.6  & -23.9  & -25.9  &  18.6  & -29.6  & -17.5  & -14.0  &  27.9  &  25.2  &  47.6 \\
AP1roG-MP2   & -1.2  &   8.7  &   3.3  &  -9.6  &  -5.7  &  -3.4  &   2.4  &   4.9  &  -3.2  &   1.7  &  -0.8  &  -0.8  &   4.4  &   1.9  &   0.1  &   0.2  &   4.4  &   3.5  &   9.6 \\
PT2SDd(d)    &  1.5  &   2.6  &   1.4  &  -5.6  &  -5.9  &  -2.8  &   5.4  &   4.6  &  -3.6  &   0.9  &   0.8  &   0.3  &   3.0  &   1.5  &   0.5  &   0.3  &   3.3  &   2.7  &   5.9 \\
PT2MDd(d)    &  2.3  &  -2.4  &  -0.6  &  -3.4  &  -6.5  &  -2.5  &   8.1  &   4.5  &  -4.4  &   0.0  &   2.1  &   1.2  &   2.3  &   1.1  &   0.6  &   0.2  &   3.6  &   2.8  &   8.1 \\
PT2MDd(sd)   &  2.3  &  -2.1  &   0.0  &  -3.7  &  -6.3  &  -2.9  &   8.2  &   4.2  &  -4.8  &  -0.1  &   1.3  &   1.2  &   2.5  &   0.9  &   0.2  &   0.1  &   3.6  &   2.7  &   8.2 \\
AP1roG-LCCD  &  0.0  &   4.0  &   1.5  &   1.3  &   2.9  &   1.3  &  -2.9  &  -2.0  &   3.7  &   0.4  &  -0.1  &  -0.2  &  -2.2  &   2.2  &   0.9  &   0.7  &   2.1  &   1.7  &   4.0 \\
AP1roG-LCCSD & -1.7  &   0.1  &  -0.2  &  -0.4  &  -1.5  &   0.6  &  -0.6  &  -0.3  &   1.1  &   1.1  &   0.1  &   0.0  &  -1.3  &   1.8  &   1.0  &   0.0  &   1.0  &   0.8  &   1.8 \\
MP2          &  5.9  &   0.9  &   2.6  &  -2.7  &  -7.1  &  -2.7  &   2.4  &   4.7  &  -0.8  &   0.6  &   0.7  &   0.4  &   2.9  &   1.7  &   0.8  &   0.7  &   3.1  &   2.5  &   7.1 \\
BCCD         &  1.4  &   0.8  &   0.4  &  -0.7  &   4.0  &   0.4  &   0.0  &  -2.0  &  -0.8  &  -3.7  &   0.3  &   0.2  &  -1.2  &  -0.2  &   0.4  &   0.0  &   1.6  &   1.1  &   4.0 \\
CCSD         &  1.2  &   0.2  &   0.3  &  -1.0  &   3.2  &   0.2  &   0.4  &  -1.6  &  -1.2  &  -3.6  &   0.3  &   0.2  &  -0.9  &  -0.4  &   0.4  &  -0.1  &   1.4  &   1.0  &   3.6 \\
CCSD(T)      & -1.7  &  -2.5  &  -1.7  &  -2.7  &  -2.5  &  -0.6  &   1.7  &   0.7  &  -2.8  &  -0.8  &   0.6  &   0.4  &   0.4  &   0.3  &   0.7  &  -0.7  &   1.6  &   1.3  &   2.8 \\
LCCD         & -2.9  &   0.3  &  -1.1  &   0.0  &   2.2  &   0.8  &  -0.8  &  -2.1  &   1.0  &  -1.2  &   0.4  &   0.3  &  -1.8  &   0.9  &   0.5  &  -0.2  &   1.4  &   1.1  &   2.9 \\
LCCSD        & -5.8  &  -4.6  &  -2.9  &  -2.1  &  -3.9  &   0.2  &   1.4  &   0.5  &  -2.0  &  -0.4  &   0.5  &   0.4  &  -0.6  &   0.9  &   0.6  &  -1.2  &   2.5  &   1.8  &   5.8 \\
\hline
\hline
\end{tabular}
}
\end{center}
\end{table*} 

In contrast to the optimized natural AP1roG basis, all PT models yield similar reaction energies with a mean absolute error of approximately 3 kcal/mol and a RMSE of about 3.5 kcal/mol, irrespective of the excitation operator $\hat{T}$ (see Table~\ref{tab:err2} and Figure~\ref{fig:err2}).
Only the combination of MP2 (for open-shell configurations) and AP1roG (for closed-shell configurations) performs worse resulting in an RMSE of 4.4 kcal/mol.
As observed above, the LCC correction outperforms all PT models if both single and double excitations are included in the cluster operator (cf.~Table~\ref{tab:err2} and Figure~\ref{fig:err2}), where the RMSE reduces to 1.0 kcal/mol.
Furthermore, the accuracy of the PT methods does not deteriorate for increasing sizes of the basis set (see Figure~\ref{fig:err2}).

Finally, we should note that the performance of all dynamic correlation corrections depends on the molecular orbital basis.
The errors with respect to CR-CCSD(T) reference data significantly decrease for PT methods with a single determinant as dual $\langle 0 \vert$ when canonical Hartree--Fock orbitals are used as molecular orbital basis.
In cases of $\langle \mathrm{AP1roG} \vert$, the dependence on the molecular orbital basis is less severe.
While reaction energies obtained by PT2MDd/PT2MDo in the canonical Hartree--Fock basis slightly improve reducing the error by a factor of 2, the mean absolute error and RMSE of PT2b remain almost unchanged compared to the natural AP1roG basis (cf.~a RMSE of 3.0 kcal/mol for AP1roG orbitals and of 3.6 kcal/mol for canonical Hartree--Fock orbitals).
Similar to PT2MD-type methods, the LCC corrections are less sensitive to the choice of the molecular orbital basis and yield reaction energies with slightly smaller error measures (cf.~a RMSE of 1.5 kcal/mol for AP1roG orbitals and LCCSD and of 1.0 kcal/mol for canonical Hartree--Fock orbitals and LCCSD).
We should emphasize that AP1roG-LCCSD in the canonical Hartree--Fock basis outperforms all conventional CC methods investigated in this work (BCC, CCSD, CCSD(T), LCCD, and LCCSD) and provides reaction energies that are within chemical accuracy compared to CR-CCSD(T) reference data.

\section{Conclusions}\label{sec:conclusion}
Wavefunctions restricted to electron-pair states allow us to reliably model static/nondynamic electron correlation effects.~\cite{Boguslawski2016}
However, in order to predict spectroscopic constants and thermochemistry within chemical accuracy, we need to include dynamic correlation effects that go beyond the simple electron-pair model.
This can be achieved \textit{a posteriori} using PT approaches,~\cite{Piotrus_PT2,Limacher_2015} coupled cluster corrections,~\cite{frozen-pCCD,Kasia-lcc} or DFT-type methods.~\cite{Garza2015,Garza-pccp}
In this work, we have extended the previously presented PT models with an AP1roG reference function and benchmarked those models against spectroscopic constants for multiply bonded diatomics and thermochemical data extrapolated to the basis set limit.
Most importantly, combining AP1roG with the investigated corrections allows us to reliably model molecular systems dominated by both static/nondynamic and dynamic electron correlation.

Specifically, our PT extensions combine a diagonal and off-diagonal one-electron zero-order Hamiltonian, a single-determinant and multi-determinant dual state, and a projection manifold restricted to double as well as single and double excitations.
In general, the performance of all PT methods can be divided in three different groups: (i) those with a diagonal zero-order Hamiltonian (PT2SDd/PT2MDd), (ii) those with an off-diagonal zero-order Hamiltonian (PT2SDo/PT2MDo), and (iii) those with an off-diagonal zero-order Hamiltonian and the full quantum-chemical Hamiltonian as perturbation operator (PT2b-type methods).
For the dissociation of multiply bonded diatomics, the PT corrections using a single-determinant dual state outperform all other investigated PT models.
In particular, the PT2SDd and PT2SDo methods (employing a diagonal and an off-diagonal zero-order Hamiltonian) provide accurate equilibrium bond lengths and vibrational frequencies compared to MRCI-SD/MRCI-SD+Q reference data, followed by AP1roG-LCCSD that is similar in accuracy, though equilibrium bond distances deviate more from reference values.
The reliable prediction of dissociation energies, however, remains challenging for all proposed PT models with errors between 15 to 30 kcal/mol.
While PT2SDd and PT2SDo provide smooth potential energy surfaces, all remaining PT methods diverge in the vicinity of dissociation.
In order to model the dissociation of multiply bonded diatomics, an LCCSD correction has to be applied.

In case of reaction energies, the accuracy of the PT corrections with respect to CR-CCSD(T) reference data increases when going from PT methods (i) to (iii) reducing the RMSE from approximately 14 kcal/mol in PT2SDd/PT2MDd, to about 5 kcal/mol in PT2SDo/PT2MDo, to 3 kcal/mol in PT2b-type methods.
The choice of the dual state and the inclusion of single excitations in the excitation operator do not significantly affect the accuracy of the PT methods (mean error, root mean square error, mean absolute error, maximum absolute error).
Furthermore, excluding pair-excited determinants from the projection manifold in PT2b-type methods improves the accuracy of PT2b only marginally.
Since pair-excitations are already described in the AP1roG reference function, it might, however, be advantageous to exclude pair excitations from the excitation operator $\hat{T}$ and hence eliminate the coupling to pair excitations modeled in the AP1roG reference function and pair excitations of the PT method, which both couple to the remaining PT amplitudes in the PT amplitude equations.

If the optimized natural AP1roG orbitals are replaced by canonical Hartree--Fock orbitals, only two distinct PT models persist, namely, PT2SDd (with double excitations) and PT2MDd (with double as well as single and double excitations).
In contrast to the natural AP1roG orbitals basis, all PT methods yield similar error measures in the canonical Hartree--Fock basis with a standard error of 3.6 kcal/mol.
Therefore, the optimization of the molecular orbital basis and the AP1roG reference determinant might be unnecessary if the molecular system is dominated by dynamic correlation and molecular properties around the equilibrium geometry are considered, provided dynamic correlation effects are accounted for in the AP1roG model.
If the optimal natural AP1roG orbitals are used in calculations, PT2b-type methods result in the smallest error measures (around 3.0 kcal/mol) and thus outperform all other PT models.
If the orbital optimization step is omitted, PT2SDd/PT2SDo provide the smallest errors that are similar to PT2b-type methods in the optimized AP1roG basis (RMSE around 3.3 kcal/mol).

Finally, a linearized coupled cluster correction with an AP1roG reference function, as presented in ref.~\citenum{Kasia-lcc}, predicts reaction energies that deviate least from CR-CCSD(T) reference data reducing the RMSE to 1.5 kcal/mol.
To minimize the error in AP1roG-LCC, single excitations are indispensable and have to be included in the cluster operator, both using optimized AP1roG orbitals and canonical Hartree--Fock orbitals.
Most importantly, the AP1roG orbital basis does not need to be optimized if chemical accuracy (approximately 1 kcal/mol) is desired for predicting equilibrium properties of weakly-correlated systems.
To conclude, AP1roG-LCCSD provides the most accurate reaction energies with respect to CR-CCSD(T) reference data, outperforming all investigated PT models as well as conventional electronic structure methods like MP2, BCC, CCSD, CCSD(T), LCCD, and LCCSD. 
In order to describe equilibrium properties of the multiply bonded diatomics C$_2$, N$_2$, and BN, PT2SDd and PT2SDo (with double excitations only) outperform AP1roG-LCCSD, fail, however, in predicting accurate dissociation energies.
Furthermore, PT2SDd/PT2SDo in the canonical Hartree--Fock basis provides the smallest errors among all investigated PT corrections (similar to MP2) and allows us to cheaply model the thermochemistry of main group elements ($\mathcal{O}(o^2v^2)$).
For strongly-correlated systems, however, the molecular orbital basis needs to be optimized before a PT2SDd or PT2SDo correction is applied.
Rotating the orbital basis increases the computational cost due to the four-index transformation and some additional prefactor of the orbital optimization.

\section{Acknowledgement}
K.B.~acknowledges financial support from a SONATA BIS grant of the National Science Centre, Poland (no.~2015/18/E/ST4/00584), a Marie-Sk\l{}odowska-Curie Individual Fellowship project no.~702635--PCCDX, and the Foundation for Polish Science for a START2016 stipend.
P.T.~thanks an OPUS grant of the National Science Centre, Poland, no.~DEC-2013/11/B/ST4/00771 and an OPUS grant of the National Science Centre, Poland, UMO-2015/19/B/ST4/02707. 

Calculations have been carried out using resources provided by Wroclaw Centre for Networking and Supercomputing (http://wcss.pl), grant no.~411 and grant no.~412.
\bibliography{rsc}
\end{document}